\documentclass[
    aps,
    prb,
    reprint,
    superscriptaddress
]{revtex4-2}
\usepackage{amsmath}
\usepackage{amssymb}
\usepackage{graphicx}
\usepackage{physics}
\usepackage{siunitx}
\usepackage{orcidlink}
\usepackage{xspace}
\usepackage{flafter}
\usepackage[version=4]{mhchem}

\newcommand{\PT}{\ensuremath{\mathcal{PT}}\xspace}
\newcommand{\tT}{\ensuremath{\vb{t}\mathcal{T}}\xspace}
\newcommand{\CT}{\ensuremath{[C_2||C_4]}\xspace}
\newcommand{\Hk}{\ensuremath{\mathcal{H}_{\vb{k}}}\xspace}

\AtBeginDocument{\RenewCommandCopy\qty\SI}

\DeclareSIUnit\angstrom{\text {Å}}

\ExplSyntaxOn
    \msg_redirect_name:nnn{siunitx}{physics-pkg}{none}
\ExplSyntaxOff

\begin{document}
\title{Strong Altermagnetic Spin Splitting and How It Arises from Nonmagnetic Atoms}

\author{Jonathan P. Mock \orcidlink{0009-0007-8268-3691}}
\email{jonathanmock@tamu.edu}
\affiliation{Department of Physics, Texas A\&M University, College Station, TX 77843-4242, USA}
\author{Wells Hero \orcidlink{0000-0003-0084-6172}}
\affiliation{Department of Physics, Texas A\&M University, College Station, TX 77843-4242, USA}
\author{Colin Lange \orcidlink{0009-0001-0661-7998}}
\affiliation{Institute of Physics, Johannes Gutenberg Universit\"at, 55099 Mainz, Germany}
\author{Rodrigo Jaeschke-Ubiergo \orcidlink{0000-0002-4821-8303}}
\affiliation{Institute of Physics, Johannes Gutenberg Universit\"at, 55099 Mainz, Germany}
\author{Jairo Sinova \orcidlink{0000-0002-9490-2333}}
\affiliation{Institute of Physics, Johannes Gutenberg Universit\"at, 55099 Mainz, Germany}
\affiliation{Department of Physics, Texas A\&M University, College Station, TX 77843-4242, USA}
\author{Artem Abanov \orcidlink{0000-0003-2322-7577}}
\affiliation{Department of Physics, Texas A\&M University, College Station, TX 77843-4242, USA}
\date{November 29, 2026}
\begin{abstract}
    Altermagnets  exhibit simultaneously zero net magnetization and nonrelativistic spin splitting, allowed by their altermagnetic crystal spin symmetry.
However, the strength of the altermagnetic spin splitting (AMSS) is determined by the specific material properties.
Here, we provide an understanding of and guidance on how to identify the relevant spin-dependent hoppings that give rise to AMSS.
We demonstrate this by identifying a minimal model with altermagnetic crystal spin symmetry, which counterintuitively gives a fully spin-degenerate band structure.
This provides the starting point to identify the relevant sources of AMSS by enabling various hopping channels.
A visual understanding of this result is that, when the hopping paths connecting sites within a spin sublattice have the nonmagnetic crystal symmetry, they do not contribute to AMSS.
Conversely, when these hopping paths are related between sublattices by the altermagnetic compensating spin symmetries, they give rise to AMSS. 
This insight allows us to create a systematic algorithm which identifies these paths from the crystal structure through the Green's function expansion of the tight-binding Hamiltonian.
The algorithm recovers the paths identified visually in our models and agrees with previous findings on specific materials.
We propose this algorithm as a tool to identify material classes that can lead to large AMSS.

\end{abstract}

\maketitle

\section{Introduction}
The discovery of altermagnetism as a distinct third phase of nonrelativistic collinear magnetism allowed for the systematic identification of materials which exhibit both zero net magnetization and strong exchange-driven spin splitting of their electronic bands~\cite{Smejkal2021a}.
Altermagnets combine the time-reversal symmetry breaking of ferromagnets~\cite{Chappert2007} with the absence of stray fields~\cite{Duine2018} and terahertz spin dynamics~\cite{Schlauderer2019,Kimel2020} of antiferromagnets~\cite{Smejkal2022a,Naka2019,Naka2021,Ma2021}.
Some of their characteristic responses, such as the anomalous Hall effect, do not scale directly with the nonrelativistic spin splitting of their bands~\cite{Smejkal2020,Smejkal2022AHEReview}.
Spin transport, however, is carried directly by this altermagnetic spin splitting (AMSS), which produces strongly anisotropic spin-polarized currents~\cite{Naka2019,Gonzalez-Hernandez2021,Smejkal2022GMR}.
For spin transport, the value of a candidate altermagnet therefore depends on how strongly it exhibits AMSS, not merely on whether AMSS is allowed.

This criterion exposes a present challenge in research on altermagnetic materials: altermagnetic crystal spin symmetries allow AMSS to manifest, but they do not determine its strength.
For example, first-principles calculations predict a maximum splitting of nearly \qty{2}{\eV} in \ce{CrSb}, roughly \qty{0.9}{\eV} in \ce{MnTe}, and \qty{0.3}{\eV} in \ce{MnF_2}~\cite{Sufyan2026,Yuan2020}, but only on the order of \qty{10}{\meV} in \ce{La_2CuO_4}~\cite{Smejkal2021a,Smejkal2022a}.
Experiments confirm the strong end of this range, with angle-resolved photoemission spectroscopy (ARPES) resolving splittings of up to \qty{1}{\eV} in \ce{CrSb}~\cite{Ding2024,Yang2024} and several hundred \unit{\meV} in \ce{MnTe}~\cite{Krempasky2024,Lee2024,Osumi2024}.
This spread persists even among altermagnets with the same type of spin-momentum locking: \ce{MnF_2} and \ce{La_2CuO_4} are both planar $d$-wave altermagnets, yet their predicted splittings differ by more than an order of magnitude~\cite{Yuan2020,Smejkal2021a,Smejkal2022a}.

The strength of the AMSS allowed by spin symmetry is set by the microscopic interactions of each material, so systematic guidance for identifying materials with potentially strong AMSS is important.
Recent studies have sought to identify these interactions using tight-binding models of real materials benchmarked against density functional theory (DFT) band structures~\cite{Mandal2025a,Camerano2026a}.
For example, models of \ce{NiS} revealed that the splitting manifests when hopping between the nonmagnetic sulfur atoms is allowed to ``modulate'' the nearest-neighbor \ce{Ni}--\ce{S} hopping~\cite{Mandal2025a}.
Similarly, models of \ce{Co_{1/4}NbSe_2} identified ``ligand-assisted coupling'' as the strongest contributor to the splitting~\cite{Camerano2026a}.
In the language of chemical bonding, both studies reveal the same insight: some hopping channels matter much more than others.
This raises the question of how to identify the hopping channels which most strongly affect AMSS, as doing so would not only clarify why some altermagnets spin-split more weakly than others, but also identify which hoppings one might tune in a real material to maximize AMSS.

To that end, we first identify a minimal model with altermagnetic crystal spin symmetry which is nonetheless fully spin degenerate.
From this zero limit, we then enable specific hopping channels to construct three variations of this model and compare their band structures with that of the spin-degenerate original.
Because every hopping respects the crystal symmetry, the comparison must be made between hopping paths, which connect sites within a spin sublattice through sequences of hops.
When such paths have the nonmagnetic crystal symmetry, they do not contribute to AMSS, but when they are related between the sublattices only by the compensating symmetry, which interchanges the opposite-spin sublattices, they give rise to it.
By visual inspection, we identify examples of these \textit{AMSS paths} in the three variations and demonstrate the lackthereof in the original model.
Mathematically, these AMSS paths correspond to terms in the power series expansion of the Green's function built from the tight-binding Hamiltonian.
Using this expansion, we formalize the comparison as an algorithm which identifies these paths directly from the crystal structure of altermagnets with one magnetic atom of each spin per primitive cell.
We check this algorithm by applying it to our models and to an $s$-orbital model of \ce{NiS}.

This path-counting algorithm determines all AMSS paths from the magnetic crystal structure which have $m\leq m_{\max}$ steps, where $m_{\max}$ is chosen by the user.
Each of these paths is listed as a symbolic expression with symmetry-derived hopping amplitudes and the corresponding spin-splitting form factor.
To first-order, these paths are mathematically equivalent to the strength of the spin-splitting arising from each path, so this allows one to identify specific chemical bonds which, when put under strain or otherwise perturbed, can maximize AMSS.

\begin{figure}[t]
    \centering
    \includegraphics{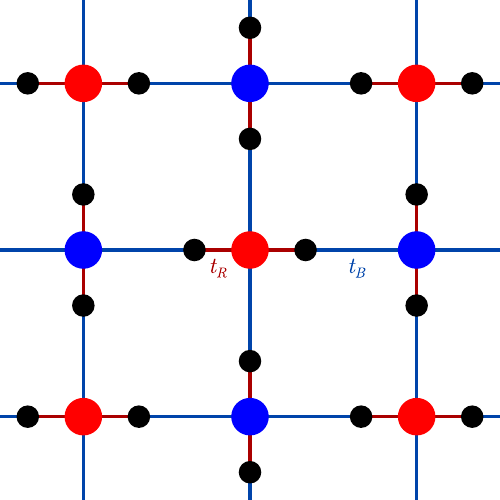}
    \caption{\label{fig:mm:model}Model altermagnet with zero AMSS.
    Large red and blue circles are magnetic atoms with opposite spins, and small black circles are nonmagnetic atoms.
    Each magnetic atom is connected to its nearest black atoms by $t_R$ (red lines) and to its next-nearest black atoms by $t_B$ (blue lines).}
\end{figure}

\begin{figure}[t]
    \centering
    \includegraphics{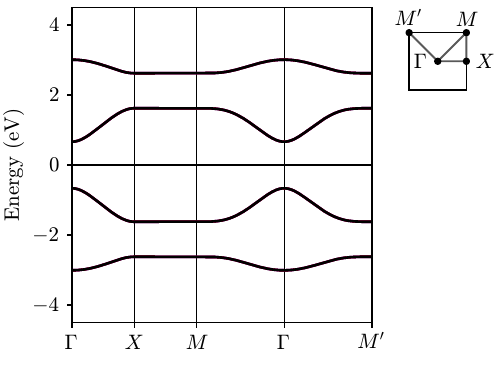}
    \caption{\label{fig:mm:bands0}Band structure of the model in Fig.~\ref*{fig:mm:model}, along the path shown in the inset, computed with $t_R=-1.25$, $t_B=-0.75$, and $J=1.0$, all in eV, and with zero on-site energy on the black atoms.
    Red and blue curves are the spin-up and spin-down bands, respectively, and coincide everywhere.
    }
\end{figure}

\begin{figure*}[t]
    \centering
    \begin{minipage}[t]{0.32\linewidth}
        (a)\\
        \includegraphics{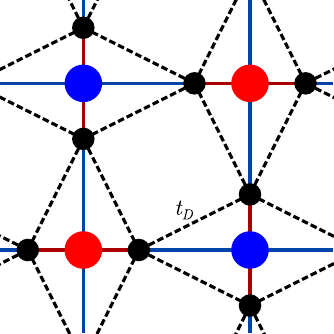}
    \end{minipage}\hfill
    \begin{minipage}[t]{0.32\linewidth}
        (b)\\
        \includegraphics{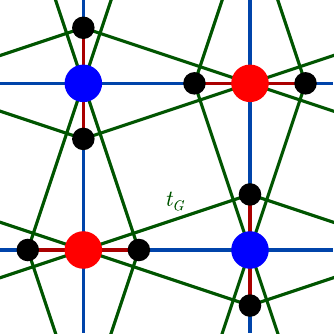}
    \end{minipage}\hfill
    \begin{minipage}[t]{0.32\linewidth}
        (c)\\
        \includegraphics{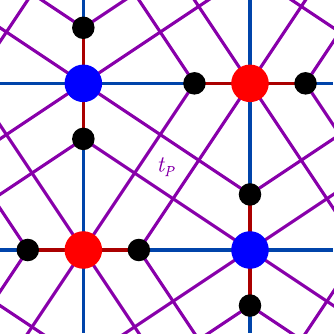}
    \end{minipage}
    \caption{\label{fig:mm:var}Variations of the model in Fig.~\ref*{fig:mm:model} which produce AMSS.
    (a) Additional hopping $t_D$ (dashed lines) between neighboring black atoms.
    (b) Additional hopping $t_G$ (green lines) between each magnetic atom and its third-nearest black atoms.
    (c) Additional hopping $t_P$ (purple lines) between each magnetic atom and its fourth-nearest black atoms.}
\end{figure*}

\begin{figure*}[t]
    \centering
    \setlength{\unitlength}{1pt}%
    \begin{picture}(504,12)
        \put(106.3,4){\makebox(0,0)[b]{(a)}}
        \put(264.8,4){\makebox(0,0)[b]{(b)}}
        \put(423.0,4){\makebox(0,0)[b]{(c)}}
    \end{picture}\\
    \includegraphics{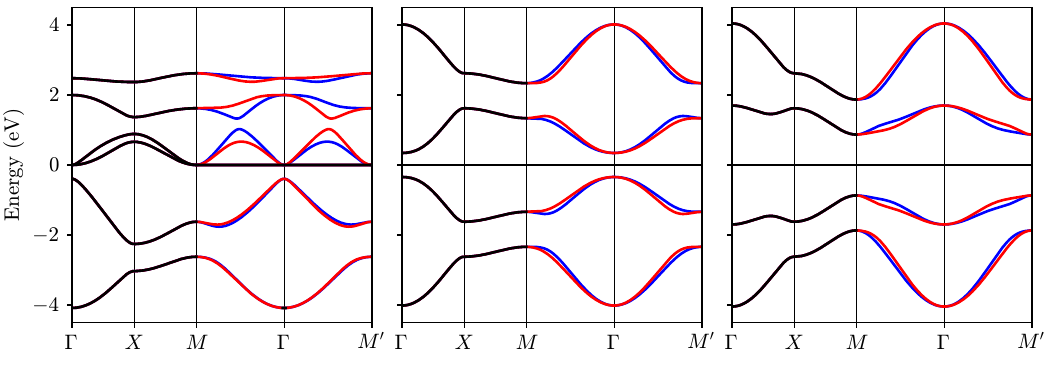}
    \caption{\label{fig:mm:bands}Band structures of the variations in Fig.~\ref*{fig:mm:var}, with (a) $t_D=-0.5$, (b) $t_G=-0.375$, and (c) $t_P=-0.375$, along the path shown in the inset of Fig.~\ref*{fig:mm:bands0}.
    All other parameters are as in Fig.~\ref*{fig:mm:bands0}, all in eV.
    Red and blue curves are the spin-up and spin-down bands, respectively.}
\end{figure*}

\section{Model Altermagnets}\label{sec:mm}

To construct a model altermagnet with zero AMSS, we recall the symmetries that define altermagnetism~\cite{Smejkal2021a,Smejkal2022a}.
First, altermagnets possess a compensating spin symmetry\,---\,a symmetry that interchanges the opposite-spin sublattices, thereby enforcing zero net magnetization.
Spin symmetries are transformations which act with generally distinct operations on spin space and real space~\cite{Brinkman1966,Litvin1974,Litvin1977}.
Such transformations are denoted $[R_i||R_j]$, where $R_i$ acts only on spin space and $R_j$ acts only on real space.
For altermagnets, the compensating spin symmetry must involve a rotation, whether proper or improper, symmorphic or nonsymmorphic.
This ensures unconventional spontaneous time-reversal symmetry breaking with $d$-, $g$-, or $i$-wave symmetry in the electronic band structure~\cite{Smejkal2021a,Smejkal2022a}.

This contrasts with conventional collinear antiferromagnets, for which at least one of the compensating symmetries must be \PT or \tT, where \PT combining time reversal ($\mathcal{T}$) and parity ($\mathcal{P}$) enforces a spin-degenerate electronic band structure, and \tT, combining $\mathcal{T}$ with translation ($\vb{t}$), enforces a nonrelativistic spin-degenerate electronic band structure.
Because we consider only collinear magnetism, \PT and \tT can be identified with $[C_2||\mathcal{P}]$ and $[C_2||\vb{t}]$, with $C_2$ replacing $\mathcal{T}$ to flip the spin.
Second, altermagnets possess neither $[C_2||\mathcal{P}]$ nor $[C_2||\vb{t}]$ and are therefore characterized by a compensating symmetry involving a rotation, such as $[C_2||C_4]$.

Guided by these two requirements, we construct the model altermagnet shown in Fig.~\ref{fig:mm:model}.
We label the magnetic atoms \textit{red} and \textit{blue} to distinguish their spins, and we label the nonmagnetic atoms \textit{black}.
Each magnetic atom is connected to its nearest black atoms by $t_R$ and to its next-nearest black atoms by $t_B$.
Because this model possesses $[C_2||C_4]$, which is a compensating symmetry, it satisfies the first requirement for being an altermagnet.
Choosing $t_R\neq t_B$ breaks both $[C_2||\mathcal{P}]$ and $[C_2||\vb{t}]$, which meets the second requirement, so the model in Fig.~\ref{fig:mm:model} is an altermagnet.
Because the real-space part of the compensating symmetry is $C_4$, we might expect this model altermagnet to exhibit $d$-wave spin splitting~\cite{Smejkal2021a}.
However, we verified both analytically (Appendix~\ref{app:zero}) and numerically (Fig.~\ref{fig:mm:bands0}) that the band structure is fully spin degenerate.

Before continuing, we emphasize that the existence of this model does not imply that real altermagnets with zero AMSS exist.
Models omit many interactions which can produce spin splitting, and in this case the most obvious omission is hopping beyond $t_R$ and $t_B$.
Tight-binding models derived from DFT generically have nonzero amplitudes for every bond allowed by symmetry.
As these amplitudes decay with distance~\cite{Marzari2012}, a realistic model should include every bond whose amplitude is not negligible.
We omit these bonds from Fig.~\ref{fig:mm:model} deliberately, to illustrate a theoretical limit rather than a physical one.

Returning to Fig.~\ref{fig:mm:model}, we consider three variations of it, each of which adds bonds that split the bands.
The first variation is shown in Fig.~\ref{fig:mm:var}(a).
Here, bonds with amplitude $t_D$ also connect neighboring black atoms.
The second and third variations are shown in Figs.~\ref{fig:mm:var}(b) and (c).
Instead of connecting black atoms to each other, we connect each magnetic atom to its third-nearest black atoms with amplitude $t_G$ or to its fourth-nearest black atoms with amplitude $t_P$, respectively.
As shown in Fig.~\ref{fig:mm:bands}, all three variations exhibit the $d$-wave AMSS expected from their \CT symmetry.
The two flat bands at zero energy in Fig.~\ref{fig:mm:bands0} and Figs.~\ref{fig:mm:bands}(b) and (c), which become dispersive in Fig.~\ref{fig:mm:bands}(a), follow from the bipartite structure of the models without $t_D$, in which no bonds connect black atoms to each other~\cite{Calugaru2022} (Appendix~\ref{app:zero}).

Two questions arise naturally.
First, why does AMSS manifest in the variations of Fig.~\ref{fig:mm:var} but not in Fig.~\ref{fig:mm:model}?
Because all four models share the same altermagnetic crystal spin symmetry, the inclusion of $t_D$, $t_G$, or $t_P$ must change something less obvious that remains consistent with the symmetry.
Second, which hopping parameters control the strength of the splitting?
Although we could scan the hopping amplitudes of these models numerically, we seek a rule that applies to any altermagnet, not just this specific model.

To answer these questions, we first define a \textit{hopping path}.
A hopping path is the sequence of hops an electron takes from one site to another, which typically crosses several bonds with different amplitudes.
Each path has an effective amplitude, given by the product of the hopping amplitudes along it, and a direction, given by its net displacement.
Since each hop contributes its amplitude times a plane wave to the Bloch Hamiltonian, the contribution of a path depends only on this product and this displacement, not on the order in which the bonds are crossed.

With this definition in mind, let $g$ be a rotation which interchanges the two spin sublattices, such as $C_4$ in our models.
In the altermagnet, $g$ is combined with the spin flip to give the compensating symmetry $[C_2||g]$, while in the nonmagnetic crystal it appears alone, as $[E||g]$, where $E$ is the identity operation.
The compensating symmetry maps each hopping path of one spin sublattice onto a path of the opposite spin sublattice with the same amplitude along the rotated direction.
If, within the same spin sublattice, a hopping path is accompanied by a path of the same amplitude along the direction rotated by $g$, we say the two paths are related by the nonmagnetic crystal symmetry $[E||g]$.

We propose that paths related by $[E||g]$ do not contribute to AMSS.
Conversely, paths which are related only by $[C_2||g]$ give rise to AMSS, and we call them \textit{AMSS paths}.
In the first case, each path has a counterpart of equal amplitude along the same direction in the other sublattice, so the two sublattices are equivalent.
In the second case, the counterpart lies along a different direction, so the two sublattices favor different directions.
This also points toward an answer to the second question, since the hoppings which control the splitting must be those along the AMSS paths.
We now apply this criterion to each model.

\subsection{Counting Paths}\label{sec:mm:paths}
Table~\ref{tab:mm:paths} lists the paths which connect a red atom to a neighboring red atom (\textit{red--red} paths) or a blue atom to a neighboring blue atom (\textit{blue--blue} paths) in Figs.~\ref{fig:mm:model} and \ref{fig:mm:var}, together with the direction each one takes.
Paths Z1--Z3 are present in Fig.~\ref{fig:mm:model} and its variations (Fig.~\ref{fig:mm:var}), while $t_D$ enables paths A1--A5, $t_G$ enables paths B1 and B2, and $t_P$ enables paths C1 and C2.
These red--red and blue--blue paths are the only candidates for AMSS paths.
Because each primitive cell of our models contains one red and one blue atom, an AMSS path appears as a difference between the red--red and blue--blue amplitudes along some direction.

\begin{table}[htb]
    \caption{\label{tab:mm:paths}Paths connecting neighboring same-color atoms in Figs.~\ref{fig:mm:model} and \ref{fig:mm:var}, up to four steps for Figs.~\ref{fig:mm:model} and \ref{fig:mm:var}(a) and up to two steps for Figs.~\ref{fig:mm:var}(b) and (c).
    Paths labeled Z are common to all models, while those labeled A, B, and C appear only with the hoppings of Figs.~\ref{fig:mm:var}(a), (b), and (c), respectively.
    \textit{Route} lists the steps of each path in order, \textit{Amplitude} gives its effective amplitude, and \textit{Red--Red} and \textit{Blue--Blue} give the direction of the neighbor it connects on each sublattice.}
    \renewcommand{\arraystretch}{1.2}
    \begin{ruledtabular}
        \begin{tabular}{ccccc}
            Path & Route & Amplitude & Red--Red & Blue--Blue \\
            \hline
            Z1 & $t_Rt_Bt_Rt_B$ & $t_R^2t_B^2$ & Diagonal & Diagonal \\
            Z2 & $t_Rt_Bt_Bt_R$ & $t_R^2t_B^2$ & Horizontal & Vertical \\
            Z3 & $t_Bt_Rt_Rt_B$ & $t_R^2t_B^2$ & Vertical & Horizontal \\
            \hline
            A1 & $t_Rt_Dt_B$ & $t_Rt_Bt_D$ & Diagonal & Diagonal \\
            A2 & $t_Rt_Dt_Dt_R$ & $t_R^2t_D^2$ & Diagonal & Diagonal \\
            A3 & $t_Bt_Dt_Dt_B$ & $t_B^2t_D^2$ & Diagonal & Diagonal \\
            A4 & $t_Rt_Dt_Dt_R$ & $t_R^2t_D^2$ & Horizontal & Vertical \\
            A5 & $t_Bt_Dt_Dt_B$ & $t_B^2t_D^2$ & Vertical & Horizontal \\
            \hline
            B1 & $t_Bt_G$ & $t_Bt_G$ & Diagonal & Diagonal \\
            B2 & $t_Gt_G$ & $t_G^2$ & Horizontal & Vertical \\
            \hline
            C1 & $t_Rt_P$ & $t_Rt_P$ & Diagonal & Diagonal \\
            C2 & $t_Pt_P$ & $t_P^2$ & Vertical & Horizontal \\
        \end{tabular}
    \end{ruledtabular}
\end{table}

Consider first the zero-AMSS model of Fig.~\ref{fig:mm:model}.
Along the diagonals, path Z1 denotes both the red--red and the blue--blue paths, with amplitude $t_R^2t_B^2$.
Path Z2 denotes the horizontal red--red and vertical blue--blue paths, and path Z3 denotes the vertical red--red and horizontal blue--blue paths.
Although the order of the hops differs between them, Z2 and Z3 have the same amplitude $t_R^2t_B^2$, so they are related by $[E||C_4]$ and are not AMSS paths.
The red--red and blue--blue paths therefore have equal amplitudes in every direction, so an electron has no preference to travel in a given direction in either sublattice, and no spin splitting occurs.

Next, consider the variation of Fig.~\ref{fig:mm:var}(a), in which $t_D$ enables paths A1--A5 in addition to Z1--Z3.
Along the diagonals, paths A1, A2, and A3 each denote both the red--red and the blue--blue paths, so the diagonal paths remain equal on the two sublattices.
Path A4 denotes the horizontal red--red and vertical blue--blue paths, with amplitude $t_R^2t_D^2$, and path A5 denotes the vertical red--red and horizontal blue--blue paths, with amplitude $t_B^2t_D^2$.
Because $|t_R|\neq|t_B|$, A4 and A5 have different amplitudes and are not related by $[E||C_4]$.
They are related only by $[C_2||C_4]$, so they are AMSS paths, consistent with the spin-split bands in Fig.~\ref{fig:mm:bands}(a).

Finally, consider the variations of Figs.~\ref{fig:mm:var}(b) and (c), in which $t_G$ enables paths B1 and B2 and $t_P$ enables paths C1 and C2.
Along the diagonals, paths B1 and C1 each denote both the red--red and the blue--blue paths.
Path B2 denotes the horizontal red--red and vertical blue--blue paths, with amplitude $t_G^2$, and no path of the same amplitude runs along the rotated direction within either sublattice.
B2 is therefore related only by $[C_2||C_4]$ and is an AMSS path, consistent with the spin-split bands in Fig.~\ref{fig:mm:bands}(b).
Likewise, path C2 denotes the vertical red--red and horizontal blue--blue paths, with amplitude $t_P^2$, and is an AMSS path, consistent with Fig.~\ref{fig:mm:bands}(c).
Unlike the paths of Figs.~\ref{fig:mm:model} and \ref{fig:mm:var}(a), B2 and C2 involve neither $t_R$ nor $t_B$.
If $t_G$ and $t_P$ were instead present together with $|t_G|=|t_P|$, as for the values used in Fig.~\ref{fig:mm:bands}, B2 and C2 would be related by $[E||C_4]$, and AMSS would vanish.

We can now answer the two questions posed earlier.
First, AMSS manifests in the variations of Fig.~\ref{fig:mm:var} but not in Fig.~\ref{fig:mm:model} because, despite its altermagnetic symmetry, the zero-AMSS model has no AMSS paths, while $t_D$, $t_G$, and $t_P$ each enable them.
Second, the hoppings which control the splitting are those along the AMSS paths.
In Fig.~\ref{fig:mm:var}(a), the AMSS paths A4 and A5 have amplitudes $t_R^2t_D^2$ and $t_B^2t_D^2$, so the splitting is proportional to $t_D^2(t_R^2-t_B^2)$ (Appendix~\ref{app:algomath}).
In Figs.~\ref{fig:mm:var}(b) and (c), the AMSS paths B2 and C2 have amplitudes $t_G^2$ and $t_P^2$, so the splitting is proportional to $t_G^2$ and $t_P^2$, respectively.

Path counting therefore identifies which hoppings control the splitting and how the splitting due to each path scales with its parameters.
It does not, however, determine how much each path contributes relative to the others, which depends on the full Hamiltonian, including where the bands lie and where they nearly cross.
For example, for the parameters of Fig.~\ref{fig:mm:bands}, the maximum splitting is \qty{0.42}{\eV} in Fig.~\ref{fig:mm:var}(a) but only \qty{0.15}{\eV} and \qty{0.18}{\eV} in Figs.~\ref{fig:mm:var}(b) and (c), even though the AMSS paths of Fig.~\ref{fig:mm:var}(a) have four steps and those of Figs.~\ref{fig:mm:var}(b) and (c) only two.

\section{Path-Counting Algorithm for AMSS}\label{sec:alg}

We now formalize the identification of AMSS paths as an algorithm.
Each path identified in Sec.~\ref{sec:mm} is one term of a moment of the Bloch Hamiltonian, a sum over all closed paths with a given number of steps~\cite{Ducastelle1970}.
The moments are the coefficients of the expansion of the Green's function in powers of the Hamiltonian, so they determine the band energies completely~\cite{Haydock1972}.
Comparing the moments between the two spin sublattices identifies every AMSS path (Appendix~\ref{app:algomath}).
Keeping the hopping amplitudes as free symbols further shows how the splitting due to each path scales with them, and therefore which bonds one would modify, through external stimuli or chemical substitution, to tune AMSS.

In the nonrelativistic limit, spin is a good quantum number for a collinear magnet, so the Bloch Hamiltonian for spin $\sigma=\pm1$ is
\begin{equation}\label{eq:alg:ham}
    \mathcal{H}^\sigma_{\vb{k}} = \mathcal{H}^0_{\vb{k}} + \sigma J\tau^z,
    \qquad
    \left(\mathcal{H}^0_{\vb{k}}\right)_{ij}=\sum_{\vb{R}}t_{ij}(\vb{R})\,e^{i\vb{k}\cdot\vb{R}},
\end{equation}

\noindent where $t_{ij}(\vb{R})$ is the amplitude for hopping from atom $i$ in the home cell to atom $j$ in the cell at lattice vector $\vb{R}$, and $\tau^z$ is diagonal with entries $s_i$, equal to $+1$ on red atoms, $-1$ on blue atoms, and $0$ on black atoms.
To first order in $J$, the $(m+1)$th moments of the two spin blocks differ by
\begin{gather}
    \Tr\left[(\Hk^\uparrow)^{m+1}-(\Hk^\downarrow)^{m+1}\right]=2(m+1)J\,A_m(\vb{k}),\label{eq:alg:diff}\\
    A_m(\vb{k})=\Tr\left[\tau^z\left(\Hk^0\right)^m\right],\label{eq:alg:Am}
\end{gather}

\noindent so $A_m$ is the difference between the $m$th moments of the red and blue atoms.
Paths with the nonmagnetic crystal symmetry contribute equally to both and cancel, so only the AMSS paths survive (Appendix~\ref{app:algomath}).

We restrict the algorithm to primitive cells with one red and one blue atom, for which this criterion is exact to all orders in $J$ (Appendix~\ref{app:algomath}), and discuss possible generalizations in Appendix~\ref{app:general}.
With this in mind, we construct the algorithm:
\begin{enumerate}
    \item Identify the primitive cell and label each atom as red, blue, or black, corresponding to $s_i=+1,-1$, and $0$, respectively.
    This cell must contain one red atom and one blue atom.
    \item Identify the symmetry group $\vb{G}$ of the nonmagnetic crystal and its subgroup $\vb{H}$ of operations which preserve each spin sublattice~\cite{Smejkal2021a}.
    The compensating symmetries are then $[C_2||g]$ with $g$ in the coset $\vb{G}-\vb{H}$.
    \item Choose a maximum bond length $r_c$, and enumerate all bonds shorter than $r_c$, assigning each bond $i$ an amplitude $t_i$.
    \item Use $\vb{G}$ to determine which bonds must have equal amplitudes: if an operation of $\vb{G}$ maps bond $j$ onto bond $i$, replace $t_j$ with $t_i$.
    \item Construct $\Hk^0$ from the remaining amplitudes, keeping each as a free symbol.
    \item Choose a maximum path length $m_{\max}$ and compute $A_m$ for $m\le m_{\max}$.
    The surviving terms come from the AMSS paths, and the first nonzero $A_m$ gives the shortest of them.
\end{enumerate}

In practice, $\vb{G}$ and $\vb{H}$ can be read off from the spin space group computed by \texttt{spinspg}~\cite{Shinohara2024}, which is built on \texttt{spglib}~\cite{Togo2024}.
To prove that a model has no AMSS, it suffices to take $m_{\max}=N-1$, where $N$ is the number of atoms in the primitive cell~\cite{SM}.

\subsection{Application to the Model Altermagnets}
We now apply our implementation of the algorithm~\cite{SM} to the models of Figs.~\ref{fig:mm:model} and \ref{fig:mm:var}.
With $r_c=0.9a$, where $a$ is the distance between neighboring same-spin magnetic atoms, Steps~3 and 4 find nine distinct bonds.
The models of Figs.~\ref{fig:mm:model} and \ref{fig:mm:var} use five of them, $t_R$, $t_B$, $t_D$, $t_G$, and $t_P$, with lengths $0.283a$, $0.424a$, $0.510a$, $0.762a$, and $0.825a$, respectively.
The primitive cell contains $N=6$ atoms, so $m_{\max}=5$ is sufficient to prove the absence of AMSS.

For Fig.~\ref{fig:mm:model}, $A_m$ vanishes identically for every $m\le m_{\max}$, so the model has no AMSS, in agreement with the fully spin-degenerate bands in Fig.~\ref{fig:mm:bands0}.
The shortest path from a red atom to the next red atom, horizontally or vertically, has amplitude $t_R^2t_B^2$, and the same is true starting from blue, so these paths cancel in Eq.~\eqref{eq:alg:Am}.

For Fig.~\ref{fig:mm:var}(a), the first nonzero $A_m$ comes from paths of $m=4$ steps,
\begin{equation}\label{eq:alg:Ab}
    A_4(\vb{k}) = 8t_D^2\left(t_B^2-t_R^2\right)\sin k_1\sin k_2,
\end{equation}

\noindent where $k_{1,2}=\vb{k}\cdot\vb{a}_{1,2}$ for the primitive lattice vectors $\vb{a}_{1,2}$ of Fig.~\ref{fig:zero:unit}, which point along the diagonals of Fig.~\ref{fig:mm:model} (Appendix~\ref{app:zero}).
Moving horizontally, the paths starting on red sum to $t_R^2t_B^2+2t_R^2t_D^2$, while those starting on blue sum to $t_R^2t_B^2+2t_B^2t_D^2$, and the two are exchanged when moving vertically.
The first term comes from paths Z2 and Z3 of Table~\ref{tab:mm:paths} and cancels.
The second comes from paths A4 and A5: $t_R$ once, $t_D$ twice, and $t_R$ once more for red, with $t_B$ replacing $t_R$ for blue.
The contribution vanishes if $t_D$ is removed or if $|t_R|=|t_B|$, and for $t_R=t_B$ the model also loses its altermagnetic symmetry.

For Figs.~\ref{fig:mm:var}(b) and (c), the first nonzero $A_m$ comes from paths of $m=2$ steps,
\begin{equation}\label{eq:alg:Ac}
    A_2(\vb{k}) = \mp8t_{G,P}^2\sin k_1\sin k_2,
\end{equation}

\noindent with the upper sign and $t_G$ for Fig.~\ref{fig:mm:var}(b) and the lower sign and $t_P$ for Fig.~\ref{fig:mm:var}(c).
With $t_G$, the paths starting on red sum to $2t_G^2$ horizontally and those starting on blue sum to $2t_G^2$ vertically, while $t_P$ exchanges these directions, which reverses the sign.
These are paths B2 and C2 of Table~\ref{tab:mm:paths}, and each contribution is proportional to the square of $t_G$ or $t_P$, as anticipated in Sec.~\ref{sec:mm:paths}.

All three variations give the same $d$-wave form factor, up to sign.
Since $\vb{a}_1\pm\vb{a}_2$ point horizontally and vertically in Fig.~\ref{fig:mm:model},
\begin{equation}\label{eq:alg:ff}
    \sin k_1\sin k_2=\frac{1}{2}\left[\cos\vb{k}\cdot(\vb{a}_1-\vb{a}_2)-\cos\vb{k}\cdot(\vb{a}_1+\vb{a}_2)\right]
\end{equation}

\noindent is half the difference between the vertical and horizontal cosine harmonics.
The splitting is therefore largest along $\Gamma$--M and $\Gamma$--M$'$ and vanishes along $\Gamma$--X and X--M.

The algorithm thus reproduces the paths found by visual inspection in Sec.~\ref{sec:mm}, and it also gives the momentum dependence of the splitting and how each contribution scales with the hopping amplitudes.

\subsection{Application to NiS}\label{sec:alg:nis}
We now apply the algorithm to \ce{NiS} in its NiAs-type structure ($P6_3/mmc$), with Ni on the $2a$ and S on the $2c$ Wyckoff positions and the experimental lattice parameters $a=\qty{3.437}{\angstrom}$ and $c=\qty{5.345}{\angstrom}$ used in Ref.~\cite{Mandal2025a}.
The magnetic order is $A$-type, consisting of ferromagnetic Ni planes stacked antiparallel along $c$.
The primitive cell contains two Ni and two S atoms, so $N=4$, and $m_{\max}=3$ is sufficient to prove the absence of AMSS.
Up to \qty{4.4}{\angstrom}, the symmetry orbits found in Step~4 contain every hopping channel of the tight-binding model in Ref.~\cite{Mandal2025a}, plus one additional Ni--S bond, as listed in Table~\ref{tab:alg:nis}.

\begin{table}[htb]
    \caption{\label{tab:alg:nis}Symmetry-distinct bonds in \ce{NiS} up to \qty{4.4}{\angstrom} and the hopping channels of Ref.~\cite{Mandal2025a} to which they correspond. Ni and Ni$'$ (S and S$'$) denote the two Ni (S) atoms of the primitive cell. The Ni--S bond at \qty{4.188}{\angstrom} is not part of the model in Ref.~\cite{Mandal2025a}.}
    \begin{ruledtabular}
        \begin{tabular}{lll}
            Bond & Length (\si{\angstrom}) & Amplitude \\
            \hline
            Ni--S & 2.392 & $t_\text{NiS}$ \\
            Ni--Ni$'$ (along $c$) & 2.673 & $t_{\text{NiNi}'}$ \\
            S--S$'$ & 3.329 & $t_{\text{SS}'}$ \\
            Ni--Ni, Ni$'$--Ni$'$ (in plane) & 3.437 & $t_\text{NiNi}$ \\
            S--S, S$'$--S$'$ (in plane) & 3.437 & $t_\text{SS}$ \\
            Ni--S & 4.188 & --- \\
            Ni--Ni$'$ & 4.354 & $t_{\text{NiNi}'}^{(2)}$ \\
        \end{tabular}
    \end{ruledtabular}
\end{table}

As constructed, the algorithm assumes one $s$-like orbital per atom, since $\Hk^0$ is indexed by atoms.
Mandal \textit{et al.} identified three cases~\cite{Mandal2025a}.
In their first two cases, each atom carries a single orbital, so the algorithm can be applied to them directly.
In Case I, only the nearest-neighbor Ni--S hopping is present, and no AMSS manifests.
The algorithm agrees: $A_m$ vanishes identically for $m\le m_{\max}$.
It continues to vanish when $t_\text{NiNi}$, $t_\text{SS}$, and both Ni--Ni$'$ amplitudes are added, which confirms their observation that these hoppings do not lift the degeneracy.
In Case II, the S--S$'$ hopping is added, and AMSS manifests.
The first nonzero $A_m$ comes from paths of $m=3$ steps,
\begin{equation}\label{eq:alg:Anis}
\begin{split}
    A_3(\vb{k}) ={}& 4t_\text{NiS}^2t_{\text{SS}'}\sin(\vb{k}\cdot\vb{c})
    \left[\sin\vb{k}\cdot(2\vb{a}_1+\vb{a}_2)\right.\\
    &\left.-\sin\vb{k}\cdot(\vb{a}_1+2\vb{a}_2)-\sin\vb{k}\cdot(\vb{a}_1-\vb{a}_2)\right],
\end{split}
\end{equation}

\noindent which vanishes if either $t_\text{NiS}$ or $t_{\text{SS}'}$ is removed, while all other amplitudes may be removed without changing this term.
This result is linear in $t_{\text{SS}'}$ and quadratic in $t_\text{NiS}$, which matches the term that the S--S$'$ hopping adds to the characteristic equation of Ref.~\cite{Mandal2025a}.
The path responsible is Ni $\to$ S $\to$ S$'$ $\to$ Ni, which ends on an equivalent Ni atom in the nearest layer of the same spin, displaced by $\pm\vb{c}$ and in plane by one of the six vectors of length $\sqrt{3}a$.

The form factor of Eq.~\eqref{eq:alg:Anis} also reproduces the nodal structure reported in Ref.~\cite{Mandal2025a}.
The factor $\sin(\vb{k}\cdot\vb{c})$ vanishes on the $k_z=0$ plane and on the zone boundary at $k_z=\pi/c$.
The bracket vanishes on the three vertical planes containing $\Gamma$--K, so this term has six in-plane sectors of alternating sign for each sign of $k_z$, giving twelve antinodal regions.
Consequently, the conventional path $\Gamma$--M--K--$\Gamma$--A--L--H--A lies entirely on nodal planes or the zone boundary and remains degenerate, while the path L$'$--$\Gamma$--L is split with opposite signs on either side of $\Gamma$.

In their third case, Mandal \textit{et al.} found that AMSS manifests without S--S$'$ hopping, provided several $d$ orbitals on Ni and $p$ orbitals on S participate in the nearest-neighbor bonding~\cite{Mandal2025a}.
This case lies outside the $s$-orbital approximation, since an $s$ orbital carries no directional information: the amplitude of a bond depends only on which orbit it belongs to, so the nearest-neighbor network alone cannot distinguish the paths that start on red from those that start on blue.
With $p$ and $d$ orbitals, the hopping between two atoms depends on the orientation of the bond relative to the orbital lobes, so short bonds can supply this directional information.
In principle, such orbitals and their hybridizations are equivalent to distinct sites with a nematic order, which enlarges the unit cell to a size that is not known in advance (Appendix~\ref{app:general}).
Within the $s$-orbital approximation, sufficiently long bonds can instead capture their effect indirectly.
For example, a Ni--S bond of length \qty{5.641}{\angstrom} produces a two-step path Ni $\to$ S $\to$ Ni with the same form factor as Eq.~\eqref{eq:alg:Anis}.
\section{Conclusion}

We have presented a model altermagnet whose bands are fully spin degenerate, which shows that the altermagnetic spin symmetry allows AMSS but does not guarantee it in effective models.
Enabling hopping between the nonmagnetic atoms, or longer-range hopping between the magnetic and nonmagnetic atoms, splits the bands.
In both cases, AMSS appears because these hoppings create AMSS paths, which lack the symmetry of the nonmagnetic crystal and are related between the sublattices only by the compensating symmetry.
The opposite directional preferences of the two spins then give the $d$-wave form of the splitting, and the hoppings along these paths control its strength.
How much each path contributes relative to the others, however, depends on the full Hamiltonian.

To make this picture usable beyond simple 2D models, we have proposed a path-counting algorithm for crystals with one magnetic atom of each spin per primitive cell.
It compares the closed paths of each length on the two spin sublattices while keeping every hopping amplitude as a free symbol, so it identifies both the paths responsible for AMSS and how the splitting due to each scales with the hoppings along it.
Through the expansion of the Green's function, it also connects the chemical-bonding picture to minimal models of altermagnets, in which these paths act as the sublattice-dependent intrasublattice hopping $t_{z,\vb{k}}$~\cite{Roig2024}.
When no such paths exist up to $N-1$ steps, it proves that the bands are spin degenerate.
Applied to our models, the algorithm recovers exactly the paths identified by visual inspection.
Applied to an $s$-orbital model of \ce{NiS}, it reproduces the results of Ref.~\cite{Mandal2025a}: among the hopping channels considered there, nearest-neighbor Ni--S hopping alone produces no AMSS, while the addition of S--S$'$ hopping produces a $g$-wave splitting with the reported nodal planes.
In each of these examples, the paths responsible pass through the nonmagnetic atoms, in line with the ligand-assisted coupling found to dominate the splitting in \ce{Co_{1/4}NbSe_2}~\cite{Camerano2026a}.

The algorithm assumes one $s$-like orbital per atom, and its possible generalizations to $p$ and $d$ orbitals and to several magnetic atoms of each spin per primitive cell are discussed in Appendix~\ref{app:general}.
Together, the mechanism and the algorithm allow a candidate material to be inspected for the hopping paths that give rise to AMSS before its band structure is computed, and they identify which hoppings one might tune to strengthen the splitting.

   \begin{acknowledgments}
AI assisted with text editing and coding.
J.M., W.H., J.S., and Ar.A. were supported on this work by the DOE Office of Science, DE-SC0026038. C.L. was supported by Deutsche Forschungsgemeinschaft (DFG, German Research Foundation)  within the priority program SPP 2558 (Project N. 581342865) and R.J. was supported by the DFG - TRR 173 – 268565370 (project A03)
\end{acknowledgments}

\appendix
\section{Analytical Solution of the Zero-AMSS Model}\label{app:zero}

\begin{figure}[t]
    \centering
    \includegraphics{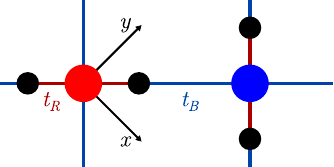}
    \caption{\label{fig:zero:unit}Primitive cell of the zero-AMSS model in Fig.~\ref*{fig:mm:model}.
    The $x$ and $y$ axes point along the primitive lattice vectors $\vb{a}_x$ and $\vb{a}_y$.}
\end{figure}

We use the primitive cell of Fig.~\ref{fig:zero:unit} and order its basis as the black atoms to the left and right of the red atom, the black atoms below and above the blue atom, and then the red and blue atoms.
We write $k_x=\vb{k}\cdot\vb{a}_x$ and $k_y=\vb{k}\cdot\vb{a}_y$, where $\vb{a}_x$ and $\vb{a}_y$ are the primitive lattice vectors $\vb{a}_1$ and $\vb{a}_2$ identified by the algorithm in Sec.~\ref{sec:alg}, so $k_x=k_1$ and $k_y=k_2$.
These vectors point along the diagonals of Fig.~\ref{fig:mm:model}, so the horizontal and vertical directions of that figure are $\vb{a}_x\pm\vb{a}_y$.
Since the black atoms have zero on-site energy and are not connected to each other, the Bloch Hamiltonian for spin $\sigma$ has the block form
\begin{equation}\label{eq:zero:ham}
    \Hk^\sigma=\begin{pmatrix}0_4&S_{\vb{k}}\\S^\dagger_{\vb{k}}&\sigma J\tau^z\end{pmatrix},
    \qquad
    S_{\vb{k}}=\begin{pmatrix}t_R&t_Be^{-i(k_x+k_y)}\\t_R&t_B\\t_Be^{ik_x}&t_R\\t_Be^{ik_y}&t_R\end{pmatrix},
\end{equation}

\noindent where $\tau^z$, restricted to the red and blue atoms, is $\operatorname{diag}(1,-1)$, and $S_{\vb{k}}$ couples the black atoms (rows) to the red and blue atoms (columns), following the convention of Eq.~\eqref{eq:alg:ham}.
Taking the Schur complement of the black block, the characteristic polynomial is
\begin{equation}\label{eq:zero:det}
    \det\left(E-\Hk^\sigma\right)=E^2\det\left(E^2-\sigma JE\tau^z-S^\dagger_{\vb{k}}S_{\vb{k}}\right),
\end{equation}

\noindent with
\begin{equation}\label{eq:zero:SS}
    S^\dagger_{\vb{k}}S_{\vb{k}}=\begin{pmatrix}2(t_R^2+t_B^2)&c_{\vb{k}}\\c^*_{\vb{k}}&2(t_R^2+t_B^2)\end{pmatrix}
\end{equation}

\noindent and $c_{\vb{k}}=t_Rt_B(1+e^{-ik_x})(1+e^{-ik_y})$.
Because the diagonal elements of $S^\dagger_{\vb{k}}S_{\vb{k}}$ are equal, $\sigma$ drops out of the determinant, and
\begin{equation}\label{eq:zero:charpoly}
\begin{split}
    \det\left(E-\Hk^\sigma\right)=E^2\big[&\left(E^2-2t_R^2-2t_B^2\right)^2-J^2E^2\\
    &-4t_R^2t_B^2(1+\cos k_x)(1+\cos k_y)\big].
\end{split}
\end{equation}

\noindent The factor $E^2$ gives two flat bands at zero energy, which consist of the two combinations of black atoms annihilated by $S^\dagger_{\vb{k}}$.
Such flat bands appear whenever the black atoms, which have zero on-site energy, outnumber the magnetic atoms and are not connected to each other~\cite{Lieb1989,Calugaru2022}.
They therefore persist when $t_G$ or $t_P$ is added, but not when $t_D$ connects the black atoms to each other.
The remaining four bands are $\pm E_\pm(\vb{k})$, with
\begin{equation}\label{eq:zero:bands}
    E_\pm^2=\gamma+\frac{J^2}{2}\pm\sqrt{\gamma J^2+\frac{J^4}{4}+|c_{\vb{k}}|^2},
\end{equation}

\noindent where $\gamma=2(t_R^2+t_B^2)$ and $|c_{\vb{k}}|^2=4t_R^2t_B^2(1+\cos k_x)(1+\cos k_y)$.
Since $J$ enters only through $J^2$, the spin-up and spin-down bands coincide at every $\vb{k}$, to all orders in $J$.

The diagonal elements of $S^\dagger_{\vb{k}}S_{\vb{k}}$ are the total squared hybridizations of the red and blue atoms with the black atoms.
They are equal because each magnetic atom has two $t_R$ bonds and two $t_B$ bonds.
In the notation of Appendix~\ref{app:algomath}, the self-energy contributed by the black atoms is $S^\dagger_{\vb{k}}S_{\vb{k}}/\omega$, so $\Delta_{\vb{k}}(\omega)$ vanishes identically.

\section{Green's Function and Moments}\label{app:algomath}

In the nonrelativistic limit, the Bloch Hamiltonian of a collinear magnet is block diagonal in spin, with the blocks $\Hk^\sigma$ of Eq.~\eqref{eq:alg:ham}.
AMSS occurs at $\vb{k}$ wherever the band energies $E^\uparrow_{\nu\vb{k}}$ and $E^\downarrow_{\nu\vb{k}}$ of the two blocks differ.
These energies are the poles of the trace of the Green's function $\mathcal{G}^\sigma_{\vb{k}}(\omega)=(\omega-\Hk^\sigma)^{-1}$,
\begin{equation}\label{eq:am:green}
    \Tr\mathcal{G}^\sigma_{\vb{k}}(\omega)=\sum_\nu\frac{1}{\omega-E^\sigma_{\nu\vb{k}}}=\sum_{m\ge0}\frac{\Tr\left[(\Hk^\sigma)^m\right]}{\omega^{m+1}},
\end{equation}

\noindent where the expansion converges for $|\omega|$ larger than every $|E^\sigma_{\nu\vb{k}}|$.
Its coefficients are the moments $\Tr\left[(\Hk^\sigma)^m\right]=\sum_\nu(E^\sigma_{\nu\vb{k}})^m$, which determine the band energies completely~\cite{Haydock1972}, so the bands are spin split at $\vb{k}$ if and only if the spin-up and spin-down moments differ for some $m$.

Because the trace does not depend on the basis, the moments can equally be evaluated in the basis of atomic states $\ket{i,\vb{R}}$.
For the real-space Hamiltonian $\mathcal{H}$, the matrix element
\begin{equation}\label{eq:am:paths}
\begin{split}
    W^{(m)}_i(\vb{R})\equiv{}&\mel{i,\vb{0}}{\mathcal{H}^m}{i,\vb{R}}\\
    ={}&\sum_{\substack{i_1,\dots,i_{m-1}\\ \vb{R}_1,\dots,\vb{R}_{m-1}}}
    t_{ii_1}(\vb{R}_1)\cdots t_{i_{m-1}i}(\vb{R}-\vb{R}_{m-1})
\end{split}
\end{equation}

\noindent is the sum over all $m$-step paths from atom $i$ to its copy in cell $\vb{R}$, each weighted by the product of the amplitudes along it~\cite{Ducastelle1970}.
Expanding the moments of $\Hk^0+\sigma J\tau^z$ to first order in $J$ gives Eq.~\eqref{eq:alg:diff}, since by the cyclicity of the trace each of the $m+1$ insertions of $\sigma J\tau^z$ contributes equally.
In the atomic basis,
\begin{equation}\label{eq:am:AmR}
    A_m(\vb{k})=\sum_{\vb{R}}e^{i\vb{k}\cdot\vb{R}}\sum_is_i\,W^{(m)}_i(\vb{R}).
\end{equation}

\noindent By the same cyclicity, each closed path is counted once for every red atom it visits and minus once for every blue atom, so a path which visits as many red atoms as blue atoms does not contribute.

Equation~\eqref{eq:am:AmR} also justifies the symmetry argument of Sec.~\ref{sec:mm}.
A compensating symmetry $[C_2||g]$ maps every red atom onto a blue atom and every path onto a path of equal amplitude, so the blue paths ending in cell $g\vb{R}$ sum to the same value as the red paths ending in cell $\vb{R}$.
If the red paths also have the nonmagnetic crystal symmetry, their sum is the same in cells $\vb{R}$ and $g\vb{R}$, so the red and blue sums are equal in every cell and cancel.
Only the AMSS paths, which are related between the sublattices by $[C_2||g]$ alone, survive.

For one red atom $r$ and one blue atom $b$ per primitive cell, the vanishing of every $A_m$ is equivalent to spin degeneracy to all orders in $J$.
We partition $\Hk^0$ into the magnetic and black atoms,
\begin{equation}\label{eq:am:part}
    \Hk^0=\begin{pmatrix}h_{\vb{k}}&V_{\vb{k}}\\V^\dagger_{\vb{k}}&H^X_{\vb{k}}\end{pmatrix},
\end{equation}

\noindent where $h_{\vb{k}}$ acts on $r$ and $b$, and $H^X_{\vb{k}}$ on the black atoms.
Downfolding the black atoms~\cite{Lowdin1951} gives the energy-dependent Hamiltonian of the magnetic atoms,
\begin{equation}\label{eq:am:heff}
    H^\text{eff}_{\vb{k}}(\omega)=h_{\vb{k}}+V_{\vb{k}}\left(\omega-H^X_{\vb{k}}\right)^{-1}V^\dagger_{\vb{k}},
\end{equation}

\noindent whose second term is the self-energy contributed by the black atoms, a sum over the paths which leave a magnetic atom, hop among the black atoms, and return.
We write $\Delta_{\vb{k}}(\omega)=a_r-a_b$ for the difference between the diagonal elements $a_r(\omega)$ and $a_b(\omega)$ of $H^\text{eff}_{\vb{k}}$, and $D_{\vb{k}}(\omega)=\det\left[\omega-H^\text{eff}_{\vb{k}}(\omega)\right]$.
The exchange enters $H^\text{eff}_{\vb{k}}$ only as $\sigma J\tau^z$ on $r$ and $b$, so
\begin{equation}\label{eq:am:charpoly}
    \det(\omega-\Hk^\sigma)=\det(\omega-H^X_{\vb{k}})\,(D_{\vb{k}}-J^2-\sigma J\Delta_{\vb{k}}),
\end{equation}

\noindent and the bands are spin degenerate at $\vb{k}$, to all orders in $J$, if and only if $\Delta_{\vb{k}}(\omega)$ vanishes identically.
At $J=0$, the trace of $\tau^z$ with the Green's function $\mathcal{G}^0_{\vb{k}}=(\omega-\Hk^0)^{-1}$ involves only the magnetic atoms, and inverting the $2\times2$ matrix $\omega-H^\text{eff}_{\vb{k}}$ gives
\begin{equation}\label{eq:am:selfenergy}
    \sum_{m\ge0}\frac{A_m(\vb{k})}{\omega^{m+1}}=\Tr\left[\tau^z\mathcal{G}^0_{\vb{k}}(\omega)\right]=\frac{\Delta_{\vb{k}}(\omega)}{D_{\vb{k}}(\omega)}.
\end{equation}

\noindent The $A_m$ therefore vanish for all $m$ if and only if $\Delta_{\vb{k}}$ does, and since $D_{\vb{k}}=\omega^2\left[1+O(1/\omega)\right]$, the leading nonzero $A_m$ is the leading term of $\Delta_{\vb{k}}$.
The Supplemental Material~\cite{SM} shows that it suffices to check $m\le N-1$, with $N$ the number of atoms in the primitive cell.
A minimal model of altermagnetism, which contains only the magnetic atoms, corresponds to such a downfolded system, whose intrasublattice hopping $t_{z,\vb{k}}$, of opposite sign on the two sublattices~\cite{Roig2024}, plays the role of $\Delta_{\vb{k}}/2$.
Equation~\eqref{eq:am:selfenergy} identifies the paths through the black atoms which generate this hopping.

These results also fix how the splitting depends on the hoppings.
To first order in $J$, the splitting of a nondegenerate band $\nu$, with eigenstate $\ket{\nu}$ of $\Hk^0$, is $\Delta E_{\nu\vb{k}}=2J\mel{\nu}{\tau^z}{\nu}$, and writing the left side of Eq.~\eqref{eq:alg:diff} in terms of the band energies gives $\sum_\nu E_{\nu\vb{k}}^m\,\Delta E_{\nu\vb{k}}=2J\,A_m(\vb{k})$, so the $A_m$ with $m\le N-1$ determine every $\Delta E_{\nu\vb{k}}$.
For Fig.~\ref{fig:mm:var}(a), each of these $A_m$ contains the factor $t_D^2(t_R^2-t_B^2)$, which comes from the AMSS paths A4 and A5, so to first order in $J$ the splitting is proportional to $t_D^2(t_R^2-t_B^2)$.
If this factor vanishes, every $A_m$ vanishes, and Eq.~\eqref{eq:am:selfenergy} then shows that the bands are spin degenerate to all orders in $J$.
The same holds for Figs.~\ref{fig:mm:var}(b) and (c) with the factors $t_G^2$ and $t_P^2$, which come from the AMSS paths B2 and C2, respectively.

\section{Possible Generalizations of the Algorithm}\label{app:general}

The algorithm of Sec.~\ref{sec:alg} assumes one $s$-like orbital per atom and one red and one blue atom per primitive cell, and we discuss here how each assumption could be relaxed.

For $s$ orbitals in the two-center approximation of Slater and Koster~\cite{Slater1954}, bonds of equal length between the same pair of species would carry equal amplitudes.
Step~4 nevertheless keeps symmetry-distinct bonds of equal length independent, since this simplification does not hold for $p$ and $d$ orbitals, whose overlap depends on the direction of the bond.

In principle, $p$ and $d$ orbitals and their hybridizations are equivalent to distinct sites with a nematic order, in which the lobes on otherwise equivalent atoms are rotated relative to one another.
This enlarges the unit cell to a size that is not known in advance.
Within the $s$-orbital approximation, sufficiently long bonds can instead capture the effect of these orbitals indirectly, as for the Ni--S bond of length \qty{5.641}{\angstrom} in Sec.~\ref{sec:alg:nis}.
A direct treatment would sum the paths over orbitals rather than atoms, with the amplitude of each bond given by the Slater--Koster two-center matrix elements~\cite{Slater1954}, whose dependence on the bond direction supplies the directional information that $s$ orbitals lack.
We expect that applying such an extension to Case III of \ce{NiS} would confirm the result of Ref.~\cite{Mandal2025a}, and we leave it to future work.

The algorithm can still be applied to cells with several magnetic sites of each spin, whether these are distinct magnetic atoms or the orbitals above, but the vanishing of $A_m$ for $m\le N-1$, with $N$ the number of sites per primitive cell, then rules out AMSS only to first order in $J$~\cite{SM}.

\bibliography{references, extra}

\begin{thebibliography}{36}%
\makeatletter
\providecommand \@ifxundefined [1]{%
 \@ifx{#1\undefined}
}%
\providecommand \@ifnum [1]{%
 \ifnum #1\expandafter \@firstoftwo
 \else \expandafter \@secondoftwo
 \fi
}%
\providecommand \@ifx [1]{%
 \ifx #1\expandafter \@firstoftwo
 \else \expandafter \@secondoftwo
 \fi
}%
\providecommand \natexlab [1]{#1}%
\providecommand \enquote  [1]{``#1''}%
\providecommand \bibnamefont  [1]{#1}%
\providecommand \bibfnamefont [1]{#1}%
\providecommand \citenamefont [1]{#1}%
\providecommand \href@noop [0]{\@secondoftwo}%
\providecommand \href [0]{\begingroup \@sanitize@url \@href}%
\providecommand \@href[1]{\@@startlink{#1}\@@href}%
\providecommand \@@href[1]{\endgroup#1\@@endlink}%
\providecommand \@sanitize@url [0]{\catcode `\\12\catcode `\$12\catcode
  `\&12\catcode `\#12\catcode `\^12\catcode `\_12\catcode `\%12\relax}%
\providecommand \@@startlink[1]{}%
\providecommand \@@endlink[0]{}%
\providecommand \url  [0]{\begingroup\@sanitize@url \@url }%
\providecommand \@url [1]{\endgroup\@href {#1}{\urlprefix }}%
\providecommand \urlprefix  [0]{URL }%
\providecommand \Eprint [0]{\href }%
\providecommand \doibase [0]{https://doi.org/}%
\providecommand \selectlanguage [0]{\@gobble}%
\providecommand \bibinfo  [0]{\@secondoftwo}%
\providecommand \bibfield  [0]{\@secondoftwo}%
\providecommand \translation [1]{[#1]}%
\providecommand \BibitemOpen [0]{}%
\providecommand \bibitemStop [0]{}%
\providecommand \bibitemNoStop [0]{.\EOS\space}%
\providecommand \EOS [0]{\spacefactor3000\relax}%
\providecommand \BibitemShut  [1]{\csname bibitem#1\endcsname}%
\let\auto@bib@innerbib\@empty
\bibitem [{\citenamefont {{\v{S}}mejkal}\ \emph
  {et~al.}(2022{\natexlab{a}})\citenamefont {{\v{S}}mejkal}, \citenamefont
  {Sinova},\ and\ \citenamefont {Jungwirth}}]{Smejkal2021a}%
  \BibitemOpen
  \bibfield  {author} {\bibinfo {author} {\bibfnamefont {L.}~\bibnamefont
  {{\v{S}}mejkal}}, \bibinfo {author} {\bibfnamefont {J.}~\bibnamefont
  {Sinova}},\ and\ \bibinfo {author} {\bibfnamefont {T.}~\bibnamefont
  {Jungwirth}},\ }\bibfield  {title} {\bibinfo {title} {{Beyond Conventional
  Ferromagnetism and Antiferromagnetism: A Phase with Nonrelativistic Spin and
  Crystal Rotation Symmetry}},\ }\href
  {https://doi.org/10.1103/PhysRevX.12.031042} {\bibfield  {journal} {\bibinfo
  {journal} {Physical Review X}\ }\textbf {\bibinfo {volume} {12}},\ \bibinfo
  {pages} {031042} (\bibinfo {year} {2022}{\natexlab{a}})}\BibitemShut
  {NoStop}%
\bibitem [{\citenamefont {Chappert}\ \emph {et~al.}(2007)\citenamefont
  {Chappert}, \citenamefont {Fert},\ and\ \citenamefont
  {Van~Dau}}]{Chappert2007}%
  \BibitemOpen
  \bibfield  {author} {\bibinfo {author} {\bibfnamefont {C.}~\bibnamefont
  {Chappert}}, \bibinfo {author} {\bibfnamefont {A.}~\bibnamefont {Fert}},\
  and\ \bibinfo {author} {\bibfnamefont {F.~N.}\ \bibnamefont {Van~Dau}},\
  }\bibfield  {title} {\bibinfo {title} {{The emergence of spin electronics in
  data storage}},\ }\href {https://doi.org/10.1038/nmat2024} {\bibfield
  {journal} {\bibinfo  {journal} {Nature Materials}\ }\textbf {\bibinfo
  {volume} {6}},\ \bibinfo {pages} {813} (\bibinfo {year} {2007})}\BibitemShut
  {NoStop}%
\bibitem [{\citenamefont {Duine}\ \emph {et~al.}(2018)\citenamefont {Duine},
  \citenamefont {Lee}, \citenamefont {Parkin},\ and\ \citenamefont
  {Stiles}}]{Duine2018}%
  \BibitemOpen
  \bibfield  {author} {\bibinfo {author} {\bibfnamefont {R.~A.}\ \bibnamefont
  {Duine}}, \bibinfo {author} {\bibfnamefont {K.-J.}\ \bibnamefont {Lee}},
  \bibinfo {author} {\bibfnamefont {S.~S.~P.}\ \bibnamefont {Parkin}},\ and\
  \bibinfo {author} {\bibfnamefont {M.~D.}\ \bibnamefont {Stiles}},\ }\bibfield
   {title} {\bibinfo {title} {{Synthetic antiferromagnetic spintronics}},\
  }\href {https://doi.org/10.1038/s41567-018-0050-y} {\bibfield  {journal}
  {\bibinfo  {journal} {Nature Physics}\ }\textbf {\bibinfo {volume} {14}},\
  \bibinfo {pages} {217} (\bibinfo {year} {2018})}\BibitemShut {NoStop}%
\bibitem [{\citenamefont {Schlauderer}\ \emph {et~al.}(2019)\citenamefont
  {Schlauderer}, \citenamefont {Lange}, \citenamefont {Baierl}, \citenamefont
  {Ebnet}, \citenamefont {Schmid}, \citenamefont {Valovcin}, \citenamefont
  {Zvezdin}, \citenamefont {Kimel}, \citenamefont {Mikhaylovskiy},\ and\
  \citenamefont {Huber}}]{Schlauderer2019}%
  \BibitemOpen
  \bibfield  {author} {\bibinfo {author} {\bibfnamefont {S.}~\bibnamefont
  {Schlauderer}}, \bibinfo {author} {\bibfnamefont {C.}~\bibnamefont {Lange}},
  \bibinfo {author} {\bibfnamefont {S.}~\bibnamefont {Baierl}}, \bibinfo
  {author} {\bibfnamefont {T.}~\bibnamefont {Ebnet}}, \bibinfo {author}
  {\bibfnamefont {C.~P.}\ \bibnamefont {Schmid}}, \bibinfo {author}
  {\bibfnamefont {D.~C.}\ \bibnamefont {Valovcin}}, \bibinfo {author}
  {\bibfnamefont {A.~K.}\ \bibnamefont {Zvezdin}}, \bibinfo {author}
  {\bibfnamefont {A.~V.}\ \bibnamefont {Kimel}}, \bibinfo {author}
  {\bibfnamefont {R.~V.}\ \bibnamefont {Mikhaylovskiy}},\ and\ \bibinfo
  {author} {\bibfnamefont {R.}~\bibnamefont {Huber}},\ }\bibfield  {title}
  {\bibinfo {title} {{Temporal and spectral fingerprints of ultrafast
  all-coherent spin switching}},\ }\href
  {https://doi.org/10.1038/s41586-019-1174-7} {\bibfield  {journal} {\bibinfo
  {journal} {Nature}\ }\textbf {\bibinfo {volume} {569}},\ \bibinfo {pages}
  {383} (\bibinfo {year} {2019})}\BibitemShut {NoStop}%
\bibitem [{\citenamefont {Kimel}\ \emph {et~al.}(2020)\citenamefont {Kimel},
  \citenamefont {Kalashnikova}, \citenamefont {Pogrebna},\ and\ \citenamefont
  {Zvezdin}}]{Kimel2020}%
  \BibitemOpen
  \bibfield  {author} {\bibinfo {author} {\bibfnamefont {A.~V.}\ \bibnamefont
  {Kimel}}, \bibinfo {author} {\bibfnamefont {A.~M.}\ \bibnamefont
  {Kalashnikova}}, \bibinfo {author} {\bibfnamefont {A.}~\bibnamefont
  {Pogrebna}},\ and\ \bibinfo {author} {\bibfnamefont {A.~K.}\ \bibnamefont
  {Zvezdin}},\ }\bibfield  {title} {\bibinfo {title} {{Fundamentals and
  perspectives of ultrafast photoferroic recording}},\ }\href
  {https://doi.org/10.1016/J.PHYSREP.2020.01.004} {\bibfield  {journal}
  {\bibinfo  {journal} {Physics Reports}\ }\textbf {\bibinfo {volume} {852}},\
  \bibinfo {pages} {1} (\bibinfo {year} {2020})}\BibitemShut {NoStop}%
\bibitem [{\citenamefont {{\v{S}}mejkal}\ \emph
  {et~al.}(2022{\natexlab{b}})\citenamefont {{\v{S}}mejkal}, \citenamefont
  {Sinova},\ and\ \citenamefont {Jungwirth}}]{Smejkal2022a}%
  \BibitemOpen
  \bibfield  {author} {\bibinfo {author} {\bibfnamefont {L.}~\bibnamefont
  {{\v{S}}mejkal}}, \bibinfo {author} {\bibfnamefont {J.}~\bibnamefont
  {Sinova}},\ and\ \bibinfo {author} {\bibfnamefont {T.}~\bibnamefont
  {Jungwirth}},\ }\bibfield  {title} {\bibinfo {title} {{Emerging Research
  Landscape of Altermagnetism}},\ }\href
  {https://doi.org/10.1103/PhysRevX.12.040501} {\bibfield  {journal} {\bibinfo
  {journal} {Physical Review X}\ }\textbf {\bibinfo {volume} {12}},\ \bibinfo
  {pages} {040501} (\bibinfo {year} {2022}{\natexlab{b}})}\BibitemShut
  {NoStop}%
\bibitem [{\citenamefont {Naka}\ \emph {et~al.}(2019)\citenamefont {Naka},
  \citenamefont {Hayami}, \citenamefont {Kusunose}, \citenamefont {Yanagi},
  \citenamefont {Motome},\ and\ \citenamefont {Seo}}]{Naka2019}%
  \BibitemOpen
  \bibfield  {author} {\bibinfo {author} {\bibfnamefont {M.}~\bibnamefont
  {Naka}}, \bibinfo {author} {\bibfnamefont {S.}~\bibnamefont {Hayami}},
  \bibinfo {author} {\bibfnamefont {H.}~\bibnamefont {Kusunose}}, \bibinfo
  {author} {\bibfnamefont {Y.}~\bibnamefont {Yanagi}}, \bibinfo {author}
  {\bibfnamefont {Y.}~\bibnamefont {Motome}},\ and\ \bibinfo {author}
  {\bibfnamefont {H.}~\bibnamefont {Seo}},\ }\bibfield  {title} {\bibinfo
  {title} {{Spin current generation in organic antiferromagnets}},\ }\href
  {https://doi.org/10.1038/s41467-019-12229-y} {\bibfield  {journal} {\bibinfo
  {journal} {Nature Communications}\ }\textbf {\bibinfo {volume} {10}},\
  \bibinfo {pages} {4305} (\bibinfo {year} {2019})}\BibitemShut {NoStop}%
\bibitem [{\citenamefont {Naka}\ \emph {et~al.}(2021)\citenamefont {Naka},
  \citenamefont {Motome},\ and\ \citenamefont {Seo}}]{Naka2021}%
  \BibitemOpen
  \bibfield  {author} {\bibinfo {author} {\bibfnamefont {M.}~\bibnamefont
  {Naka}}, \bibinfo {author} {\bibfnamefont {Y.}~\bibnamefont {Motome}},\ and\
  \bibinfo {author} {\bibfnamefont {H.}~\bibnamefont {Seo}},\ }\bibfield
  {title} {\bibinfo {title} {{Perovskite as a spin current generator}},\ }\href
  {https://doi.org/10.1103/PhysRevB.103.125114} {\bibfield  {journal} {\bibinfo
   {journal} {Physical Review B}\ }\textbf {\bibinfo {volume} {103}},\ \bibinfo
  {pages} {125114} (\bibinfo {year} {2021})}\BibitemShut {NoStop}%
\bibitem [{\citenamefont {Ma}\ \emph {et~al.}(2021)\citenamefont {Ma},
  \citenamefont {Hu}, \citenamefont {Li}, \citenamefont {Liu}, \citenamefont
  {Yao}, \citenamefont {Jia},\ and\ \citenamefont {Liu}}]{Ma2021}%
  \BibitemOpen
  \bibfield  {author} {\bibinfo {author} {\bibfnamefont {H.-Y.}\ \bibnamefont
  {Ma}}, \bibinfo {author} {\bibfnamefont {M.}~\bibnamefont {Hu}}, \bibinfo
  {author} {\bibfnamefont {N.}~\bibnamefont {Li}}, \bibinfo {author}
  {\bibfnamefont {J.}~\bibnamefont {Liu}}, \bibinfo {author} {\bibfnamefont
  {W.}~\bibnamefont {Yao}}, \bibinfo {author} {\bibfnamefont {J.-F.}\
  \bibnamefont {Jia}},\ and\ \bibinfo {author} {\bibfnamefont {J.}~\bibnamefont
  {Liu}},\ }\bibfield  {title} {\bibinfo {title} {{Multifunctional
  antiferromagnetic materials with giant piezomagnetism and noncollinear spin
  current}},\ }\href {https://doi.org/10.1038/s41467-021-23127-7} {\bibfield
  {journal} {\bibinfo  {journal} {Nature Communications}\ }\textbf {\bibinfo
  {volume} {12}},\ \bibinfo {pages} {2846} (\bibinfo {year}
  {2021})}\BibitemShut {NoStop}%
\bibitem [{\citenamefont {{\v{S}}mejkal}\ \emph {et~al.}(2020)\citenamefont
  {{\v{S}}mejkal}, \citenamefont {Gonz{\'{a}}lez-Hern{\'{a}}ndez},
  \citenamefont {Jungwirth},\ and\ \citenamefont {Sinova}}]{Smejkal2020}%
  \BibitemOpen
  \bibfield  {author} {\bibinfo {author} {\bibfnamefont {L.}~\bibnamefont
  {{\v{S}}mejkal}}, \bibinfo {author} {\bibfnamefont {R.}~\bibnamefont
  {Gonz{\'{a}}lez-Hern{\'{a}}ndez}}, \bibinfo {author} {\bibfnamefont
  {T.}~\bibnamefont {Jungwirth}},\ and\ \bibinfo {author} {\bibfnamefont
  {J.}~\bibnamefont {Sinova}},\ }\bibfield  {title} {\bibinfo {title} {{Crystal
  time-reversal symmetry breaking and spontaneous Hall effect in collinear
  antiferromagnets}},\ }\bibfield  {journal} {\bibinfo  {journal} {Science
  Advances}\ }\textbf {\bibinfo {volume} {6}},\ \href
  {https://doi.org/10.1126/sciadv.aaz8809} {10.1126/sciadv.aaz8809} (\bibinfo
  {year} {2020})\BibitemShut {NoStop}%
\bibitem [{\citenamefont {{\v{S}}mejkal}\ \emph
  {et~al.}(2022{\natexlab{c}})\citenamefont {{\v{S}}mejkal}, \citenamefont
  {MacDonald}, \citenamefont {Sinova}, \citenamefont {Nakatsuji},\ and\
  \citenamefont {Jungwirth}}]{Smejkal2022AHEReview}%
  \BibitemOpen
  \bibfield  {author} {\bibinfo {author} {\bibfnamefont {L.}~\bibnamefont
  {{\v{S}}mejkal}}, \bibinfo {author} {\bibfnamefont {A.~H.}\ \bibnamefont
  {MacDonald}}, \bibinfo {author} {\bibfnamefont {J.}~\bibnamefont {Sinova}},
  \bibinfo {author} {\bibfnamefont {S.}~\bibnamefont {Nakatsuji}},\ and\
  \bibinfo {author} {\bibfnamefont {T.}~\bibnamefont {Jungwirth}},\ }\bibfield
  {title} {\bibinfo {title} {{Anomalous Hall antiferromagnets}},\ }\href
  {https://doi.org/10.1038/s41578-022-00430-3} {\bibfield  {journal} {\bibinfo
  {journal} {Nature Reviews Materials}\ }\textbf {\bibinfo {volume} {7}},\
  \bibinfo {pages} {482} (\bibinfo {year} {2022}{\natexlab{c}})}\BibitemShut
  {NoStop}%
\bibitem [{\citenamefont {Gonz{\'{a}}lez-Hern{\'{a}}ndez}\ \emph
  {et~al.}(2021)\citenamefont {Gonz{\'{a}}lez-Hern{\'{a}}ndez}, \citenamefont
  {{\v{S}}mejkal}, \citenamefont {V{\'{y}}born{\'{y}}}, \citenamefont {Yahagi},
  \citenamefont {Sinova}, \citenamefont {Jungwirth},\ and\ \citenamefont
  {{\v{Z}}elezn{\'{y}}}}]{Gonzalez-Hernandez2021}%
  \BibitemOpen
  \bibfield  {author} {\bibinfo {author} {\bibfnamefont {R.}~\bibnamefont
  {Gonz{\'{a}}lez-Hern{\'{a}}ndez}}, \bibinfo {author} {\bibfnamefont
  {L.}~\bibnamefont {{\v{S}}mejkal}}, \bibinfo {author} {\bibfnamefont
  {K.}~\bibnamefont {V{\'{y}}born{\'{y}}}}, \bibinfo {author} {\bibfnamefont
  {Y.}~\bibnamefont {Yahagi}}, \bibinfo {author} {\bibfnamefont
  {J.}~\bibnamefont {Sinova}}, \bibinfo {author} {\bibfnamefont
  {T.}~\bibnamefont {Jungwirth}},\ and\ \bibinfo {author} {\bibfnamefont
  {J.}~\bibnamefont {{\v{Z}}elezn{\'{y}}}},\ }\bibfield  {title} {\bibinfo
  {title} {{Efficient Electrical Spin Splitter Based on Nonrelativistic
  Collinear Antiferromagnetism}},\ }\href
  {https://doi.org/10.1103/PhysRevLett.126.127701} {\bibfield  {journal}
  {\bibinfo  {journal} {Physical Review Letters}\ }\textbf {\bibinfo {volume}
  {126}},\ \bibinfo {pages} {127701} (\bibinfo {year} {2021})}\BibitemShut
  {NoStop}%
\bibitem [{\citenamefont {{\v{S}}mejkal}\ \emph
  {et~al.}(2022{\natexlab{d}})\citenamefont {{\v{S}}mejkal}, \citenamefont
  {Hellenes}, \citenamefont {Gonz{\'{a}}lez-Hern{\'{a}}ndez}, \citenamefont
  {Sinova},\ and\ \citenamefont {Jungwirth}}]{Smejkal2022GMR}%
  \BibitemOpen
  \bibfield  {author} {\bibinfo {author} {\bibfnamefont {L.}~\bibnamefont
  {{\v{S}}mejkal}}, \bibinfo {author} {\bibfnamefont {A.~B.}\ \bibnamefont
  {Hellenes}}, \bibinfo {author} {\bibfnamefont {R.}~\bibnamefont
  {Gonz{\'{a}}lez-Hern{\'{a}}ndez}}, \bibinfo {author} {\bibfnamefont
  {J.}~\bibnamefont {Sinova}},\ and\ \bibinfo {author} {\bibfnamefont
  {T.}~\bibnamefont {Jungwirth}},\ }\bibfield  {title} {\bibinfo {title}
  {{Giant and Tunneling Magnetoresistance in Unconventional Collinear
  Antiferromagnets with Nonrelativistic Spin-Momentum Coupling}},\ }\href
  {https://doi.org/10.1103/PhysRevX.12.011028} {\bibfield  {journal} {\bibinfo
  {journal} {Physical Review X}\ }\textbf {\bibinfo {volume} {12}},\ \bibinfo
  {pages} {011028} (\bibinfo {year} {2022}{\natexlab{d}})}\BibitemShut
  {NoStop}%
\bibitem [{\citenamefont {Sufyan}\ \emph {et~al.}(2026)\citenamefont {Sufyan},
  \citenamefont {Marfoua}, \citenamefont {Larsson}, \citenamefont {van Loon},\
  and\ \citenamefont {Armiento}}]{Sufyan2026}%
  \BibitemOpen
  \bibfield  {author} {\bibinfo {author} {\bibfnamefont {A.}~\bibnamefont
  {Sufyan}}, \bibinfo {author} {\bibfnamefont {B.}~\bibnamefont {Marfoua}},
  \bibinfo {author} {\bibfnamefont {J.~A.}\ \bibnamefont {Larsson}}, \bibinfo
  {author} {\bibfnamefont {E.}~\bibnamefont {van Loon}},\ and\ \bibinfo
  {author} {\bibfnamefont {R.}~\bibnamefont {Armiento}},\ }\bibfield  {title}
  {\bibinfo {title} {{High-throughput quantification of altermagnetic band
  splitting}},\ }\href {https://doi.org/10.1103/mmdm-hrj4} {\bibfield
  {journal} {\bibinfo  {journal} {Physical Review Materials}\ }\textbf
  {\bibinfo {volume} {10}},\ \bibinfo {pages} {044407} (\bibinfo {year}
  {2026})}\BibitemShut {NoStop}%
\bibitem [{\citenamefont {Yuan}\ \emph {et~al.}(2020)\citenamefont {Yuan},
  \citenamefont {Wang}, \citenamefont {Luo}, \citenamefont {Rashba},\ and\
  \citenamefont {Zunger}}]{Yuan2020}%
  \BibitemOpen
  \bibfield  {author} {\bibinfo {author} {\bibfnamefont {L.-D.}\ \bibnamefont
  {Yuan}}, \bibinfo {author} {\bibfnamefont {Z.}~\bibnamefont {Wang}}, \bibinfo
  {author} {\bibfnamefont {J.-W.}\ \bibnamefont {Luo}}, \bibinfo {author}
  {\bibfnamefont {E.~I.}\ \bibnamefont {Rashba}},\ and\ \bibinfo {author}
  {\bibfnamefont {A.}~\bibnamefont {Zunger}},\ }\bibfield  {title} {\bibinfo
  {title} {{Giant momentum-dependent spin splitting in centrosymmetric low-Z
  antiferromagnets}},\ }\href {https://doi.org/10.1103/PhysRevB.102.014422}
  {\bibfield  {journal} {\bibinfo  {journal} {Physical Review B}\ }\textbf
  {\bibinfo {volume} {102}},\ \bibinfo {pages} {014422} (\bibinfo {year}
  {2020})}\BibitemShut {NoStop}%
\bibitem [{\citenamefont {Ding}\ \emph {et~al.}(2024)\citenamefont {Ding},
  \citenamefont {Jiang}, \citenamefont {Chen}, \citenamefont {Tao},
  \citenamefont {Liu}, \citenamefont {Li}, \citenamefont {Liu}, \citenamefont
  {Sun}, \citenamefont {Cheng}, \citenamefont {Liu}, \citenamefont {Yang},
  \citenamefont {Zhang}, \citenamefont {Deng}, \citenamefont {Jing},
  \citenamefont {Huang}, \citenamefont {Shi}, \citenamefont {Ye}, \citenamefont
  {Qiao}, \citenamefont {Wang}, \citenamefont {Guo}, \citenamefont {Feng},\
  and\ \citenamefont {Shen}}]{Ding2024}%
  \BibitemOpen
  \bibfield  {author} {\bibinfo {author} {\bibfnamefont {J.}~\bibnamefont
  {Ding}}, \bibinfo {author} {\bibfnamefont {Z.}~\bibnamefont {Jiang}},
  \bibinfo {author} {\bibfnamefont {X.}~\bibnamefont {Chen}}, \bibinfo {author}
  {\bibfnamefont {Z.}~\bibnamefont {Tao}}, \bibinfo {author} {\bibfnamefont
  {Z.}~\bibnamefont {Liu}}, \bibinfo {author} {\bibfnamefont {T.}~\bibnamefont
  {Li}}, \bibinfo {author} {\bibfnamefont {J.}~\bibnamefont {Liu}}, \bibinfo
  {author} {\bibfnamefont {J.}~\bibnamefont {Sun}}, \bibinfo {author}
  {\bibfnamefont {J.}~\bibnamefont {Cheng}}, \bibinfo {author} {\bibfnamefont
  {J.}~\bibnamefont {Liu}}, \bibinfo {author} {\bibfnamefont {Y.}~\bibnamefont
  {Yang}}, \bibinfo {author} {\bibfnamefont {R.}~\bibnamefont {Zhang}},
  \bibinfo {author} {\bibfnamefont {L.}~\bibnamefont {Deng}}, \bibinfo {author}
  {\bibfnamefont {W.}~\bibnamefont {Jing}}, \bibinfo {author} {\bibfnamefont
  {Y.}~\bibnamefont {Huang}}, \bibinfo {author} {\bibfnamefont
  {Y.}~\bibnamefont {Shi}}, \bibinfo {author} {\bibfnamefont {M.}~\bibnamefont
  {Ye}}, \bibinfo {author} {\bibfnamefont {S.}~\bibnamefont {Qiao}}, \bibinfo
  {author} {\bibfnamefont {Y.}~\bibnamefont {Wang}}, \bibinfo {author}
  {\bibfnamefont {Y.}~\bibnamefont {Guo}}, \bibinfo {author} {\bibfnamefont
  {D.}~\bibnamefont {Feng}},\ and\ \bibinfo {author} {\bibfnamefont
  {D.}~\bibnamefont {Shen}},\ }\bibfield  {title} {\bibinfo {title} {{Large
  Band Splitting in g-Wave Altermagnet CrSb}},\ }\href
  {https://doi.org/10.1103/PhysRevLett.133.206401} {\bibfield  {journal}
  {\bibinfo  {journal} {Physical Review Letters}\ }\textbf {\bibinfo {volume}
  {133}},\ \bibinfo {pages} {206401} (\bibinfo {year} {2024})}\BibitemShut
  {NoStop}%
\bibitem [{\citenamefont {Yang}\ \emph {et~al.}(2025)\citenamefont {Yang},
  \citenamefont {Li}, \citenamefont {Yang}, \citenamefont {Li}, \citenamefont
  {Zheng}, \citenamefont {Zhu}, \citenamefont {Pan}, \citenamefont {Xu},
  \citenamefont {Cao}, \citenamefont {Zhao}, \citenamefont {Jana},
  \citenamefont {Zhang}, \citenamefont {Ye}, \citenamefont {Song},
  \citenamefont {Hu}, \citenamefont {Yang}, \citenamefont {Fujii},
  \citenamefont {Vobornik}, \citenamefont {Shi}, \citenamefont {Yuan},
  \citenamefont {Zhang}, \citenamefont {Xu},\ and\ \citenamefont
  {Liu}}]{Yang2024}%
  \BibitemOpen
  \bibfield  {author} {\bibinfo {author} {\bibfnamefont {G.}~\bibnamefont
  {Yang}}, \bibinfo {author} {\bibfnamefont {Z.}~\bibnamefont {Li}}, \bibinfo
  {author} {\bibfnamefont {S.}~\bibnamefont {Yang}}, \bibinfo {author}
  {\bibfnamefont {J.}~\bibnamefont {Li}}, \bibinfo {author} {\bibfnamefont
  {H.}~\bibnamefont {Zheng}}, \bibinfo {author} {\bibfnamefont
  {W.}~\bibnamefont {Zhu}}, \bibinfo {author} {\bibfnamefont {Z.}~\bibnamefont
  {Pan}}, \bibinfo {author} {\bibfnamefont {Y.}~\bibnamefont {Xu}}, \bibinfo
  {author} {\bibfnamefont {S.}~\bibnamefont {Cao}}, \bibinfo {author}
  {\bibfnamefont {W.}~\bibnamefont {Zhao}}, \bibinfo {author} {\bibfnamefont
  {A.}~\bibnamefont {Jana}}, \bibinfo {author} {\bibfnamefont {J.}~\bibnamefont
  {Zhang}}, \bibinfo {author} {\bibfnamefont {M.}~\bibnamefont {Ye}}, \bibinfo
  {author} {\bibfnamefont {Y.}~\bibnamefont {Song}}, \bibinfo {author}
  {\bibfnamefont {L.-H.}\ \bibnamefont {Hu}}, \bibinfo {author} {\bibfnamefont
  {L.}~\bibnamefont {Yang}}, \bibinfo {author} {\bibfnamefont {J.}~\bibnamefont
  {Fujii}}, \bibinfo {author} {\bibfnamefont {I.}~\bibnamefont {Vobornik}},
  \bibinfo {author} {\bibfnamefont {M.}~\bibnamefont {Shi}}, \bibinfo {author}
  {\bibfnamefont {H.}~\bibnamefont {Yuan}}, \bibinfo {author} {\bibfnamefont
  {Y.}~\bibnamefont {Zhang}}, \bibinfo {author} {\bibfnamefont
  {Y.}~\bibnamefont {Xu}},\ and\ \bibinfo {author} {\bibfnamefont
  {Y.}~\bibnamefont {Liu}},\ }\bibfield  {title} {\bibinfo {title}
  {{Three-dimensional mapping of the altermagnetic spin splitting in CrSb}},\
  }\href {https://doi.org/10.1038/s41467-025-56647-7} {\bibfield  {journal}
  {\bibinfo  {journal} {Nature Communications}\ }\textbf {\bibinfo {volume}
  {16}},\ \bibinfo {pages} {1442} (\bibinfo {year} {2025})}\BibitemShut
  {NoStop}%
\bibitem [{\citenamefont {Krempask{\'{y}}}\ \emph {et~al.}(2024)\citenamefont
  {Krempask{\'{y}}}, \citenamefont {{\v{S}}mejkal}, \citenamefont {D’Souza},
  \citenamefont {Hajlaoui}, \citenamefont {Springholz}, \citenamefont
  {Uhl{\'{i}}{\v{r}}ov{\'{a}}}, \citenamefont {Alarab}, \citenamefont
  {Constantinou}, \citenamefont {Strocov}, \citenamefont {Usanov},
  \citenamefont {Pudelko}, \citenamefont {Gonz{\'{a}}lez-Hern{\'{a}}ndez},
  \citenamefont {Birk~Hellenes}, \citenamefont {Jansa}, \citenamefont
  {Reichlov{\'{a}}}, \citenamefont {{\v{S}}ob{\'{a}}ň}, \citenamefont
  {Gonzalez~Betancourt}, \citenamefont {Wadley}, \citenamefont {Sinova},
  \citenamefont {Kriegner}, \citenamefont {Min{\'{a}}r}, \citenamefont {Dil},\
  and\ \citenamefont {Jungwirth}}]{Krempasky2024}%
  \BibitemOpen
  \bibfield  {author} {\bibinfo {author} {\bibfnamefont {J.}~\bibnamefont
  {Krempask{\'{y}}}}, \bibinfo {author} {\bibfnamefont {L.}~\bibnamefont
  {{\v{S}}mejkal}}, \bibinfo {author} {\bibfnamefont {S.~W.}\ \bibnamefont
  {D’Souza}}, \bibinfo {author} {\bibfnamefont {M.}~\bibnamefont {Hajlaoui}},
  \bibinfo {author} {\bibfnamefont {G.}~\bibnamefont {Springholz}}, \bibinfo
  {author} {\bibfnamefont {K.}~\bibnamefont {Uhl{\'{i}}{\v{r}}ov{\'{a}}}},
  \bibinfo {author} {\bibfnamefont {F.}~\bibnamefont {Alarab}}, \bibinfo
  {author} {\bibfnamefont {P.~C.}\ \bibnamefont {Constantinou}}, \bibinfo
  {author} {\bibfnamefont {V.}~\bibnamefont {Strocov}}, \bibinfo {author}
  {\bibfnamefont {D.}~\bibnamefont {Usanov}}, \bibinfo {author} {\bibfnamefont
  {W.~R.}\ \bibnamefont {Pudelko}}, \bibinfo {author} {\bibfnamefont
  {R.}~\bibnamefont {Gonz{\'{a}}lez-Hern{\'{a}}ndez}}, \bibinfo {author}
  {\bibfnamefont {A.}~\bibnamefont {Birk~Hellenes}}, \bibinfo {author}
  {\bibfnamefont {Z.}~\bibnamefont {Jansa}}, \bibinfo {author} {\bibfnamefont
  {H.}~\bibnamefont {Reichlov{\'{a}}}}, \bibinfo {author} {\bibfnamefont
  {Z.}~\bibnamefont {{\v{S}}ob{\'{a}}ň}}, \bibinfo {author} {\bibfnamefont
  {R.~D.}\ \bibnamefont {Gonzalez~Betancourt}}, \bibinfo {author}
  {\bibfnamefont {P.}~\bibnamefont {Wadley}}, \bibinfo {author} {\bibfnamefont
  {J.}~\bibnamefont {Sinova}}, \bibinfo {author} {\bibfnamefont
  {D.}~\bibnamefont {Kriegner}}, \bibinfo {author} {\bibfnamefont
  {J.}~\bibnamefont {Min{\'{a}}r}}, \bibinfo {author} {\bibfnamefont {J.~H.}\
  \bibnamefont {Dil}},\ and\ \bibinfo {author} {\bibfnamefont {T.}~\bibnamefont
  {Jungwirth}},\ }\bibfield  {title} {\bibinfo {title} {{Altermagnetic lifting
  of Kramers spin degeneracy}},\ }\href
  {https://doi.org/10.1038/s41586-023-06907-7} {\bibfield  {journal} {\bibinfo
  {journal} {Nature}\ }\textbf {\bibinfo {volume} {626}},\ \bibinfo {pages}
  {517} (\bibinfo {year} {2024})}\BibitemShut {NoStop}%
\bibitem [{\citenamefont {Lee}\ \emph {et~al.}(2024)\citenamefont {Lee},
  \citenamefont {Lee}, \citenamefont {Jung}, \citenamefont {Jung},
  \citenamefont {Kim}, \citenamefont {Lee}, \citenamefont {Seok}, \citenamefont
  {Kim}, \citenamefont {Park}, \citenamefont {{\v{S}}mejkal}, \citenamefont
  {Kang},\ and\ \citenamefont {Kim}}]{Lee2024}%
  \BibitemOpen
  \bibfield  {author} {\bibinfo {author} {\bibfnamefont {S.~S.}\ \bibnamefont
  {Lee}}, \bibinfo {author} {\bibfnamefont {S.~S.}\ \bibnamefont {Lee}},
  \bibinfo {author} {\bibfnamefont {S.}~\bibnamefont {Jung}}, \bibinfo {author}
  {\bibfnamefont {J.}~\bibnamefont {Jung}}, \bibinfo {author} {\bibfnamefont
  {D.}~\bibnamefont {Kim}}, \bibinfo {author} {\bibfnamefont {Y.}~\bibnamefont
  {Lee}}, \bibinfo {author} {\bibfnamefont {B.}~\bibnamefont {Seok}}, \bibinfo
  {author} {\bibfnamefont {J.}~\bibnamefont {Kim}}, \bibinfo {author}
  {\bibfnamefont {B.~G.}\ \bibnamefont {Park}}, \bibinfo {author}
  {\bibfnamefont {L.}~\bibnamefont {{\v{S}}mejkal}}, \bibinfo {author}
  {\bibfnamefont {C.-J.}\ \bibnamefont {Kang}},\ and\ \bibinfo {author}
  {\bibfnamefont {C.}~\bibnamefont {Kim}},\ }\bibfield  {title} {\bibinfo
  {title} {{Broken Kramers Degeneracy in Altermagnetic MnTe}},\ }\href
  {https://doi.org/10.1103/PhysRevLett.132.036702} {\bibfield  {journal}
  {\bibinfo  {journal} {Physical Review Letters}\ }\textbf {\bibinfo {volume}
  {132}},\ \bibinfo {pages} {036702} (\bibinfo {year} {2024})}\BibitemShut
  {NoStop}%
\bibitem [{\citenamefont {Osumi}\ \emph {et~al.}(2024)\citenamefont {Osumi},
  \citenamefont {Souma}, \citenamefont {Aoyama}, \citenamefont {Yamauchi},
  \citenamefont {Honma}, \citenamefont {Nakayama}, \citenamefont {Takahashi},
  \citenamefont {Ohgushi},\ and\ \citenamefont {Sato}}]{Osumi2024}%
  \BibitemOpen
  \bibfield  {author} {\bibinfo {author} {\bibfnamefont {T.}~\bibnamefont
  {Osumi}}, \bibinfo {author} {\bibfnamefont {S.}~\bibnamefont {Souma}},
  \bibinfo {author} {\bibfnamefont {T.}~\bibnamefont {Aoyama}}, \bibinfo
  {author} {\bibfnamefont {K.}~\bibnamefont {Yamauchi}}, \bibinfo {author}
  {\bibfnamefont {A.}~\bibnamefont {Honma}}, \bibinfo {author} {\bibfnamefont
  {K.}~\bibnamefont {Nakayama}}, \bibinfo {author} {\bibfnamefont
  {T.}~\bibnamefont {Takahashi}}, \bibinfo {author} {\bibfnamefont
  {K.}~\bibnamefont {Ohgushi}},\ and\ \bibinfo {author} {\bibfnamefont
  {T.}~\bibnamefont {Sato}},\ }\bibfield  {title} {\bibinfo {title}
  {{Observation of a giant band splitting in altermagnetic MnTe}},\ }\href
  {https://doi.org/10.1103/PhysRevB.109.115102} {\bibfield  {journal} {\bibinfo
   {journal} {Physical Review B}\ }\textbf {\bibinfo {volume} {109}},\ \bibinfo
  {pages} {115102} (\bibinfo {year} {2024})}\BibitemShut {NoStop}%
\bibitem [{\citenamefont {Mandal}\ \emph {et~al.}(2025)\citenamefont {Mandal},
  \citenamefont {Das},\ and\ \citenamefont {Nanda}}]{Mandal2025a}%
  \BibitemOpen
  \bibfield  {author} {\bibinfo {author} {\bibfnamefont {A.}~\bibnamefont
  {Mandal}}, \bibinfo {author} {\bibfnamefont {A.}~\bibnamefont {Das}},\ and\
  \bibinfo {author} {\bibfnamefont {B.~R.~K.}\ \bibnamefont {Nanda}},\
  }\bibfield  {title} {\bibinfo {title} {{Deterministic role of chemical
  bonding in the formation of altermagnetism: Reflection from the correlated
  electron system NiS}},\ }\href {https://doi.org/10.1103/sm63-1dcx} {\bibfield
   {journal} {\bibinfo  {journal} {Physical Review B}\ }\textbf {\bibinfo
  {volume} {112}},\ \bibinfo {pages} {014420} (\bibinfo {year}
  {2025})}\BibitemShut {NoStop}%
\bibitem [{\citenamefont {Camerano}\ \emph {et~al.}(2026)\citenamefont
  {Camerano}, \citenamefont {Bisti},\ and\ \citenamefont
  {Profeta}}]{Camerano2026a}%
  \BibitemOpen
  \bibfield  {author} {\bibinfo {author} {\bibfnamefont {L.}~\bibnamefont
  {Camerano}}, \bibinfo {author} {\bibfnamefont {F.}~\bibnamefont {Bisti}},\
  and\ \bibinfo {author} {\bibfnamefont {G.}~\bibnamefont {Profeta}},\
  }\bibfield  {title} {\bibinfo {title} {{Ligand-mediated origin of
  altermagnetic spin splitting}},\ }\href {https://doi.org/10.1103/s83n-w474}
  {\bibfield  {journal} {\bibinfo  {journal} {Physical Review B}\ }\textbf
  {\bibinfo {volume} {114}},\ \bibinfo {pages} {264408} (\bibinfo {year}
  {2026})}\BibitemShut {NoStop}%
\bibitem [{\citenamefont {Brinkman}\ and\ \citenamefont
  {Elliott}(1966)}]{Brinkman1966}%
  \BibitemOpen
  \bibfield  {author} {\bibinfo {author} {\bibfnamefont {W.~F.}\ \bibnamefont
  {Brinkman}}\ and\ \bibinfo {author} {\bibfnamefont {R.~J.}\ \bibnamefont
  {Elliott}},\ }\bibfield  {title} {\bibinfo {title} {{Theory of Spin-Space
  Groups}},\ }\href {https://doi.org/10.1098/rspa.1966.0211} {\bibfield
  {journal} {\bibinfo  {journal} {Proceedings of the Royal Society A:
  Mathematical, Physical and Engineering Sciences}\ }\textbf {\bibinfo {volume}
  {294}},\ \bibinfo {pages} {343} (\bibinfo {year} {1966})}\BibitemShut
  {NoStop}%
\bibitem [{\citenamefont {Litvin}\ and\ \citenamefont
  {Opechowski}(1974)}]{Litvin1974}%
  \BibitemOpen
  \bibfield  {author} {\bibinfo {author} {\bibfnamefont {D.~B.}\ \bibnamefont
  {Litvin}}\ and\ \bibinfo {author} {\bibfnamefont {W.}~\bibnamefont
  {Opechowski}},\ }\bibfield  {title} {\bibinfo {title} {{Spin groups}},\
  }\href {https://doi.org/10.1016/0031-8914(74)90157-8} {\bibfield  {journal}
  {\bibinfo  {journal} {Physica}\ }\textbf {\bibinfo {volume} {76}},\ \bibinfo
  {pages} {538} (\bibinfo {year} {1974})}\BibitemShut {NoStop}%
\bibitem [{\citenamefont {Litvin}(1977)}]{Litvin1977}%
  \BibitemOpen
  \bibfield  {author} {\bibinfo {author} {\bibfnamefont {D.~B.}\ \bibnamefont
  {Litvin}},\ }\bibfield  {title} {\bibinfo {title} {{Spin point groups}},\
  }\href {https://doi.org/10.1107/S0567739477000709} {\bibfield  {journal}
  {\bibinfo  {journal} {Acta Crystallographica Section A}\ }\textbf {\bibinfo
  {volume} {33}},\ \bibinfo {pages} {279} (\bibinfo {year} {1977})}\BibitemShut
  {NoStop}%
\bibitem [{\citenamefont {Marzari}\ \emph {et~al.}(2012)\citenamefont
  {Marzari}, \citenamefont {Mostofi}, \citenamefont {Yates}, \citenamefont
  {Souza},\ and\ \citenamefont {Vanderbilt}}]{Marzari2012}%
  \BibitemOpen
  \bibfield  {author} {\bibinfo {author} {\bibfnamefont {N.}~\bibnamefont
  {Marzari}}, \bibinfo {author} {\bibfnamefont {A.~A.}\ \bibnamefont
  {Mostofi}}, \bibinfo {author} {\bibfnamefont {J.~R.}\ \bibnamefont {Yates}},
  \bibinfo {author} {\bibfnamefont {I.}~\bibnamefont {Souza}},\ and\ \bibinfo
  {author} {\bibfnamefont {D.}~\bibnamefont {Vanderbilt}},\ }\bibfield  {title}
  {\bibinfo {title} {{Maximally localized Wannier functions: Theory and
  applications}},\ }\href {https://doi.org/10.1103/RevModPhys.84.1419}
  {\bibfield  {journal} {\bibinfo  {journal} {Reviews of Modern Physics}\
  }\textbf {\bibinfo {volume} {84}},\ \bibinfo {pages} {1419} (\bibinfo {year}
  {2012})}\BibitemShut {NoStop}%
\bibitem [{\citenamefont {C{\u{a}}lug{\u{a}}ru}\ \emph
  {et~al.}(2022)\citenamefont {C{\u{a}}lug{\u{a}}ru}, \citenamefont {Chew},
  \citenamefont {Elcoro}, \citenamefont {Xu}, \citenamefont {Regnault},
  \citenamefont {Song},\ and\ \citenamefont {Bernevig}}]{Calugaru2022}%
  \BibitemOpen
  \bibfield  {author} {\bibinfo {author} {\bibfnamefont {D.}~\bibnamefont
  {C{\u{a}}lug{\u{a}}ru}}, \bibinfo {author} {\bibfnamefont {A.}~\bibnamefont
  {Chew}}, \bibinfo {author} {\bibfnamefont {L.}~\bibnamefont {Elcoro}},
  \bibinfo {author} {\bibfnamefont {Y.}~\bibnamefont {Xu}}, \bibinfo {author}
  {\bibfnamefont {N.}~\bibnamefont {Regnault}}, \bibinfo {author}
  {\bibfnamefont {Z.-D.}\ \bibnamefont {Song}},\ and\ \bibinfo {author}
  {\bibfnamefont {B.~A.}\ \bibnamefont {Bernevig}},\ }\bibfield  {title}
  {\bibinfo {title} {{General construction and topological classification of
  crystalline flat bands}},\ }\href
  {https://doi.org/10.1038/s41567-021-01445-3} {\bibfield  {journal} {\bibinfo
  {journal} {Nature Physics}\ }\textbf {\bibinfo {volume} {18}},\ \bibinfo
  {pages} {185} (\bibinfo {year} {2022})}\BibitemShut {NoStop}%
\bibitem [{\citenamefont {Ducastelle}\ and\ \citenamefont
  {Cyrot-Lackmann}(1970)}]{Ducastelle1970}%
  \BibitemOpen
  \bibfield  {author} {\bibinfo {author} {\bibfnamefont {F.}~\bibnamefont
  {Ducastelle}}\ and\ \bibinfo {author} {\bibfnamefont {F.}~\bibnamefont
  {Cyrot-Lackmann}},\ }\bibfield  {title} {\bibinfo {title} {{Moments
  developments and their application to the electronic charge distribution of d
  bands}},\ }\href {https://doi.org/10.1016/0022-3697(70)90134-4} {\bibfield
  {journal} {\bibinfo  {journal} {Journal of Physics and Chemistry of Solids}\
  }\textbf {\bibinfo {volume} {31}},\ \bibinfo {pages} {1295} (\bibinfo {year}
  {1970})}\BibitemShut {NoStop}%
\bibitem [{\citenamefont {Haydock}\ \emph {et~al.}(1972)\citenamefont
  {Haydock}, \citenamefont {Heine},\ and\ \citenamefont {Kelly}}]{Haydock1972}%
  \BibitemOpen
  \bibfield  {author} {\bibinfo {author} {\bibfnamefont {R.}~\bibnamefont
  {Haydock}}, \bibinfo {author} {\bibfnamefont {V.}~\bibnamefont {Heine}},\
  and\ \bibinfo {author} {\bibfnamefont {M.~J.}\ \bibnamefont {Kelly}},\
  }\bibfield  {title} {\bibinfo {title} {{Electronic structure based on the
  local atomic environment for tight-binding bands}},\ }\href
  {https://doi.org/10.1088/0022-3719/5/20/004} {\bibfield  {journal} {\bibinfo
  {journal} {Journal of Physics C: Solid State Physics}\ }\textbf {\bibinfo
  {volume} {5}},\ \bibinfo {pages} {2845} (\bibinfo {year} {1972})}\BibitemShut
  {NoStop}%
\bibitem [{\citenamefont {Shinohara}\ \emph {et~al.}(2024)\citenamefont
  {Shinohara}, \citenamefont {Togo}, \citenamefont {Watanabe}, \citenamefont
  {Nomoto}, \citenamefont {Tanaka},\ and\ \citenamefont
  {Arita}}]{Shinohara2024}%
  \BibitemOpen
  \bibfield  {author} {\bibinfo {author} {\bibfnamefont {K.}~\bibnamefont
  {Shinohara}}, \bibinfo {author} {\bibfnamefont {A.}~\bibnamefont {Togo}},
  \bibinfo {author} {\bibfnamefont {H.}~\bibnamefont {Watanabe}}, \bibinfo
  {author} {\bibfnamefont {T.}~\bibnamefont {Nomoto}}, \bibinfo {author}
  {\bibfnamefont {I.}~\bibnamefont {Tanaka}},\ and\ \bibinfo {author}
  {\bibfnamefont {R.}~\bibnamefont {Arita}},\ }\bibfield  {title} {\bibinfo
  {title} {{Algorithm for spin symmetry operation search}},\ }\href
  {https://doi.org/10.1107/S2053273323009257} {\bibfield  {journal} {\bibinfo
  {journal} {Acta Crystallographica Section A Foundations and Advances}\
  }\textbf {\bibinfo {volume} {80}},\ \bibinfo {pages} {94} (\bibinfo {year}
  {2024})}\BibitemShut {NoStop}%
\bibitem [{\citenamefont {Togo}\ \emph {et~al.}(2024)\citenamefont {Togo},
  \citenamefont {Shinohara},\ and\ \citenamefont {Tanaka}}]{Togo2024}%
  \BibitemOpen
  \bibfield  {author} {\bibinfo {author} {\bibfnamefont {A.}~\bibnamefont
  {Togo}}, \bibinfo {author} {\bibfnamefont {K.}~\bibnamefont {Shinohara}},\
  and\ \bibinfo {author} {\bibfnamefont {I.}~\bibnamefont {Tanaka}},\
  }\bibfield  {title} {\bibinfo {title} {{Spglib: a software library for
  crystal symmetry search}},\ }\href
  {https://doi.org/10.1080/27660400.2024.2384822} {\bibfield  {journal}
  {\bibinfo  {journal} {Science and Technology of Advanced Materials: Methods}\
  }\textbf {\bibinfo {volume} {4}},\ \bibinfo {pages} {2384822} (\bibinfo
  {year} {2024})}\BibitemShut {NoStop}%
\bibitem [{SM()}]{SM}%
  \BibitemOpen
  \href@noop {} {}\bibinfo {note} {The Supplemental Material, containing a
  description of the program, its source code, the input files used in this
  work, and the proof that it suffices to check $m\le N-1$, will be made
  available with the published version of this article.}\BibitemShut {Stop}%
\bibitem [{\citenamefont {Roig}\ \emph {et~al.}(2024)\citenamefont {Roig},
  \citenamefont {Kreisel}, \citenamefont {Yu}, \citenamefont {Andersen},\ and\
  \citenamefont {Agterberg}}]{Roig2024}%
  \BibitemOpen
  \bibfield  {author} {\bibinfo {author} {\bibfnamefont {M.}~\bibnamefont
  {Roig}}, \bibinfo {author} {\bibfnamefont {A.}~\bibnamefont {Kreisel}},
  \bibinfo {author} {\bibfnamefont {Y.}~\bibnamefont {Yu}}, \bibinfo {author}
  {\bibfnamefont {B.~M.}\ \bibnamefont {Andersen}},\ and\ \bibinfo {author}
  {\bibfnamefont {D.~F.}\ \bibnamefont {Agterberg}},\ }\bibfield  {title}
  {\bibinfo {title} {{Minimal models for altermagnetism}},\ }\href
  {https://doi.org/10.1103/PhysRevB.110.144412} {\bibfield  {journal} {\bibinfo
   {journal} {Physical Review B}\ }\textbf {\bibinfo {volume} {110}},\ \bibinfo
  {pages} {144412} (\bibinfo {year} {2024})}\BibitemShut {NoStop}%
\bibitem [{\citenamefont {Lieb}(1989)}]{Lieb1989}%
  \BibitemOpen
  \bibfield  {author} {\bibinfo {author} {\bibfnamefont {E.~H.}\ \bibnamefont
  {Lieb}},\ }\bibfield  {title} {\bibinfo {title} {{Two theorems on the Hubbard
  model}},\ }\href {https://doi.org/10.1103/PhysRevLett.62.1201} {\bibfield
  {journal} {\bibinfo  {journal} {Physical Review Letters}\ }\textbf {\bibinfo
  {volume} {62}},\ \bibinfo {pages} {1201} (\bibinfo {year}
  {1989})}\BibitemShut {NoStop}%
\bibitem [{\citenamefont {L{\"{o}}wdin}(1951)}]{Lowdin1951}%
  \BibitemOpen
  \bibfield  {author} {\bibinfo {author} {\bibfnamefont {P.-O.}\ \bibnamefont
  {L{\"{o}}wdin}},\ }\bibfield  {title} {\bibinfo {title} {{A Note on the
  Quantum-Mechanical Perturbation Theory}},\ }\href
  {https://doi.org/10.1063/1.1748067} {\bibfield  {journal} {\bibinfo
  {journal} {The Journal of Chemical Physics}\ }\textbf {\bibinfo {volume}
  {19}},\ \bibinfo {pages} {1396} (\bibinfo {year} {1951})}\BibitemShut
  {NoStop}%
\bibitem [{\citenamefont {Slater}\ and\ \citenamefont
  {Koster}(1954)}]{Slater1954}%
  \BibitemOpen
  \bibfield  {author} {\bibinfo {author} {\bibfnamefont {J.~C.}\ \bibnamefont
  {Slater}}\ and\ \bibinfo {author} {\bibfnamefont {G.~F.}\ \bibnamefont
  {Koster}},\ }\bibfield  {title} {\bibinfo {title} {{Simplified LCAO method
  for the periodic potential problem}},\ }\href
  {https://doi.org/10.1103/PhysRev.94.1498} {\bibfield  {journal} {\bibinfo
  {journal} {Physical Review}\ }\textbf {\bibinfo {volume} {94}},\ \bibinfo
  {pages} {1498} (\bibinfo {year} {1954})}\BibitemShut {NoStop}%
\end{thebibliography}%

\end{document}